\documentclass{rstransa}

\newcommand{\ket}{{\rangle}}
\newcommand{\bra}{{\langle}}
\newcommand{\bs}{\boldsymbol}
\newcommand{\cA}{\mathcal{A}}
\newcommand{\cH}{\mathcal{H}}
\newcommand{\cU}{\mathcal{U}}

\begin{document}

\title{Counterdiabatic Formalism of Shortcuts to Adiabaticity}

\author{
Mikio Nakahara$^{1}$}

\address{$^{1}$Research Institute for Science and Technology\\
Kindai University, 577-8502, Japan}

\subject{shortcuts to adiabaticity, counterdiabatic formalism}

\keywords{quantum control, counterdiabatic driving, transitionless driving}

\corres{Mikio Nakahara\\
\email{nakahara@rist.kindai.ac.jp}}

\begin{abstract}
Pedagogical introduction to counterdiabatic formalism of shortcuts to adiabaticity is given so that the readers are accessible to some of more specialized articles in the rest of this theme issue without a much barrier. 
A guide to references is given so that this article also serves as a mini-review.
\end{abstract}


\begin{fmtext}
\section{Introduction}

Precise control of a quantum system in short time is indispensable to fight against decoherence and implement a large scale quantum computer. Although adiabatic quantum control is known to shuttle a system to a final destination with high precision, it takes long time to achieve high fidelity and quantum state would degrade during the process in the presence of decoherence and noise. Shortcuts to adiabaticity (STA for short) is a comprehensive approach to achieve the goal of adiabatic quantum control with much shorter time. 

Counterdiabatic (CD for short) formalism, also known as transitionless quantum driving, is an approach to STA by modifying the Hamiltonian so that the quantum state follows the adiabatic path of the original Hamiltonian to the goal in shorter time. This means that there are nonadiabatic transitions among eigenstates of the modified Hamiltonian. The formalism has been tested in many physical as well as chemical systems, a part of which will be reported in this theme issue.   

In this article, we outline aspects of the CD formalism. In the next section, we briefly introduce the formalism. In section 3,
simple examples of (i) spins driven by time-dependent magnetic fields and (ii) a harmonic oscillator with time-dependent trap frequency are outlined.
We re-derive the CD Hamiltonian from a slightly different
viewpoint in section 4.
STA based on the dynamical invariant (DI for short) is also a popular
approach to quantum control theory. 
We discuss the relationship between the CD and the DI formalisms in section 5. 
Applications and demonstrations of the CD formalism are briefly outlined in section 6 so that this introduction
\end{fmtext}

\maketitle
\noindent
serves as a mini-review and a guide for further reading.

It is impossible to exhaust all aspects of CD formalism in this article
because of limited space. Interested readers are advised to consult with excellent reviews \cite{rev1,rev2,rev3} and a special issue \cite{rev4}
for details of the subject and other approaches to STA.

\section{Counterdiabatic Formalism}

The CD formalism was first introduced in
\cite{1.emmanouilidou} in a restricted form and subsequently
formulated in more general settings in \cite{2.demirplak,2a.demirplak,3.berry}.
We closely follow \cite{3.berry} and \cite{ref:deffner,ref:adolfo,ref:jarzynski} in this section.

Let $H_0(t)$ be an arbitrary Hamiltonian acting on a finite-dimensional Hilbert space $\mathbb{C}^d$ with $0 \leq t \leq T$
and let $|n(t) \ket$ be the $n$th instantaneous eigenvector of $H_0(t)$ with eigenvalue $E_n(t)$;
\begin{equation}\label{eq:eigen}
H_0(t)|n(t) \ket = E_n(t)|n(t) \ket.
\end{equation}
It is known that an initial eigenstate remains the instantaneous eigenstate
during the time-evolution if the variation of $H_0(t)$,
and hence $|n(t) \ket$, is slow enough
compared to the energy gap so that the adiabatic condition
\begin{equation}\label{eq:adiabatic}
\hbar \left| \frac{\bra m(t)|\partial_t n(t)\ket}{E_n(t)-E_m(t)}\right|
\ll 1
\end{equation}
is satisfied \cite{4.Born,5.Kato} for $n \neq m$, where $|\partial_t n(t) \ket =
\partial_t|n(t) \ket$.

As mentioned in section 1, slow adiabatic change of a state
is a problem in view of decoherence and it is desirable to 
obtain the same result as adiabatic time-evolution in much shorter
time, which is of help to implement high precision quantum gates
in the gate model of quantum computing. Suppression of decoherence is of vital importance in the NISQ (Noisy Intermediate-Scale Quantum Computing) 
without quantum error correcting codes.
Prevention of transition to other eigenstates is also essential in adiabatic quantum computing. 
Chen \textit{et al.} \cite{ref:chen} introduced ``shortcuts
to adiabaticity'' 
for controlling atoms in a harmonic trap
by employing the method developed by Lewis and Riesenfeld \cite{lewis}, which resulted in a surge of STA research. 

In this article, we introduce the CD
approach to STA. Recall first that the solution of the 
time-dependent Schr\"odinger equation
$i \hbar \partial_t |\psi_n  (t) \ket =H_0(t)|\psi_n(t) \ket$
with the initial condition $|\psi_n(0) \ket = |n(0)\ket$ is
\begin{equation}\label{eq:adia}
|\psi_n(t) \ket = e^{i \xi_n(t)}  |n(t) \ket,\quad
\xi_n(t) = -\frac{1}{\hbar}\int_0^t E_n(s) ds + i \int_0^t \bra n(s)|\partial_s n(s) \ket ds
\end{equation}
if $H_0(t)$ satisfies the condition (\ref{eq:adiabatic}).
Here the phase of $|n(t) \ket$ has been chosen and fixed arbitrarily.
The first term of the exponent $\xi_n(t)$ is the conventional dynamical 
phase while the second term is the geometric phase \cite{berryphase}. Let $H_0(t)$ depend on time through a set of parameters, collectively denoted as ${\bs \lambda}(t)$, namely $H_0(t) = H_0({\bs \lambda}(t))$, where ${\bs \lambda}(t) \in \mathbb{R}^N$, $N$ being the number
of independent parameters. Correspondingly $|n(t) \ket$ is also 
written as $|n({\bs \lambda}(t))\ket$. The geometric phase is independent of the speed of time-evolution but only depends on the path in the parameter space so far as the time-evolution is adiabatic;
$$
\int_0^t \bra n(s) |\partial_s n(s) \ket ds =
\int_{\bs \lambda(0)}^{{\bs \lambda}(t)} \bra n(\bs \lambda)| \partial_{\mu} n(\bs \lambda) \ket d\lambda^{\mu},
$$
where $\partial_{\mu}$ stands for $\partial/\partial \lambda^{\mu}$. The one-form $\mathcal{A}=\bra n(\bs \lambda)| \partial_{\mu} n(\bs \lambda) \ket d\lambda^{\mu}$ is called the
Berry connection and plays the role of a gauge potential. 
In fact, if the phase of $|n(\bs \lambda)\ket$ is redefined as $e^{i \chi (\bs \lambda)}|n(\bs \lambda)
\ket$, then $\mathcal{A}_{\mu}$ changes as $\mathcal{A}_{\mu}
+i \partial_{\mu} \chi(\bs \lambda)$. The geometric phase
reduces to the Berry phase in case the path $\bs \lambda(t)$
is closed.

Suppose there exists a Hamiltonian $H(t)$
\textit{to be determined} such that
\begin{equation}\label{eq:sch}
i \hbar \frac{d}{dt} |\psi_n(t) \ket = H(t) |\psi_n(t) \ket
\end{equation}
for a vector $|\psi_n(t) \ket$ of the form (\ref{eq:adia}).
We usually solve $|\psi_n(t) \ket$ for a given $H(t)$ with some initial
condition. In contrast,
finding $H(t)$ for a given $|\psi_n(t) \ket$ is an inverse problem.
$H(t)$ reduces to $H_0(t)$ if the evolution is adiabatic.
However, we require here Eq.~(\ref{eq:sch})
be satisfied with possibly nonadiabatic time-evolution. Let
$|\psi_n(0) \ket = |n(0)\ket$ and 
\begin{equation}
U(t) = \sum_n|\psi_n(t) \ket \bra n(0)| 
= \sum_n \exp\left[-\frac{i}{\hbar} \int_0^t E_n(s) ds - \int_0^t \bra n(s)|
\partial_s n(s) \ket ds \right]|n(t) \ket \bra n(0) |
\end{equation}
be the time-evolution operator derived from (\ref{eq:sch}).
Then
\begin{eqnarray}
H(t) &=&i \hbar  (\partial_t U(t)) U(t)^{-1}\nonumber\\
&=& \sum_n E_n(t)|n(t) \ket \bra n(t)|+
 i \hbar\sum_n (I - |n(t) \ket \bra n(t)|) |\partial_t n(t) \ket \bra n(t)|
 \label{eq:hamcd}
\end{eqnarray}
Here the first term is nothing but $H_0(t)$ while the
second term
\begin{equation}\label{eq:h1}
H_1(t) =  i \hbar \sum_n (I - |n(t) \ket \bra n(t)|) |\partial_t n(t) \ket \bra n(t)| =i \hbar 
\sum_{n,\mu} \dot{\lambda}^{\mu} (I - |n(t) \ket \bra n(t)|) |\partial_{\mu} n(t) \ket \bra n(t)| 
\end{equation}
is called the CD term. Nonadiabatic time-evolution would
make the wave function $|\psi_n(t) \ket$ deviate from
the adiabatic path of $H_0(t)$ but 
$H_1(t)$ pushes the path back to $|\psi_n(t) \ket$.
It is natural, in view of this, that $H_1(t)$ is sizable when
 $\bs{\lambda}$ change rapidly.
Note also that $H_1(t)$ is independent of the state to be driven. $H_1(t)$ is also written as $ i\hbar \sum_n \dot{P}_n(t) P_n(t)$, where $P_n(t)=|n(t) \ket \bra n(t)|$.

Suppose one drives a car on an icy road.
The driver will slow down on a curve to avoid slipping out the
road if the road is flat. However, the driver can keep the same 
speed if the road is banked. $H_1(t)$ plays the role of the bank
in this analogy. The bank is designed precisely if the curvature 
radius of the road and the speed of the car are specified.
This is what $H_1(t)$ will do to keep the path $|\psi_n(t) \ket$ transitionless, $|\psi_n (t) \ket \propto |n(t) \ket$, 
in the original time-evolution of $H_0(t)$.
Since $H_0(t)$ and $H(t)$ do not commute in general,
the time-evolution of $|\psi_n(t) \ket$ cannot be transitionless with respect to $H(t)$.

Several remarks are in order.
\begin{itemize}
\item $H_1(t)$ is expressed in a slightly more compact form if the 
completeness relation of $\{|n(t) \ket\}$ is employed as
\begin{equation}\label{eq:abc}
H_1(t) =i\hbar \sum_{m \neq n} |m(t) \ket\bra m(t)|\partial_t n(t) \ket
\bra n(t)|.
\end{equation}  
It is clear that $H_1(t)$ has vanishing diagonal elements 
with respect to $\{|n(t) \ket\}$ basis.
\item
For some applications, it is convenient
to write $H_1(t)$ as
\begin{equation}\label{eq:newh}
H_1(t) = i\hbar \sum_{m \neq n}  
\frac{\bra m(t)|\partial_t H_0(t)|n(t) \ket }
{E_n(t) - E_m(t) }|m(t) \ket\bra n(t)|,
\end{equation}
where we assume there are no degenerate eigenvalues.
\footnote{See \cite{ktaka} for cases with degenerate eigenvalues.}
Equation (\ref{eq:newh}) is proved by taking $t$-derivative of (\ref{eq:eigen}) and 
multiply $\bra m(t)|$ from the left to obtain
$$
\bra m (t)|\partial_t n(t) \ket = \frac{\bra m(t)|\partial_t H_0(t)|
n(t) \ket}{E_n(t)-E_m(t)} \qquad (m \neq n)
$$
then substituting this into (\ref{eq:abc}).

\item Two terms $H_0(t)$ and $H_1(t)$ are orthogonal,
\begin{equation}\label{frobe}
\bra H_0(t), H_1(t) \ket = 0,
\end{equation}
where we employed the Frobenius inner product, $
\bra A, B \ket = \mathrm{tr}~(A^{\dagger} B)$. A Hamiltonian is a generator of the time-evolution and an element of the
Lie algebra, living in the tangent space of the Lie group U($d$).
(\ref{frobe}) tells us that the time-evolution due to $H_1(t)$ is
always orthogonal to that generated by $H_0(t)$.  
$H_1(t)$ is also orthogonal to $\dot{H}_0(t)$ as
\begin{equation}
\bra \dot{H}_0(t), H_1(t) \ket =
i\hbar \sum_{m \neq n} (E_n-E_m)|\bra 
m|\dot{n} \ket|^2 =0,
\end{equation}
where we noted that the left-hand side is a real number.

\item It was mentioned in the beginning of this section that each
eigenvector $|n (t) \ket$ of $H_0(t)$ has a gauge degree of
freedom, which means $H_1(t)$ is not uniquely determined
\cite{DRx}.
This freedom can be used to find the optimal $H_1(t)$ for
a given requirement, such as minimum intensity of the control field
for example. 
\end{itemize}

We have considered so far a quantum system whose Hamiltonian is represented by a matrix of a finite dimension. 
Next we consider a system whose Hamiltonian takes
a form of a differential operator. For this purposes, it is desirable
to express $H_1(t)$ also in terms of a differential operator
rather than that in the form (\ref{eq:h1}). 

We analyze a class of Hamiltonians with a ``scale-invariant''
property
\begin{equation}\label{eq:generalinv}
H_0(\bs \lambda(t)) = \frac{p^2}{2m}+ U[q, \bs \lambda(t)]=
\frac{p^2}{2m} + \frac{1}{\gamma(t)^2} U_0 \left[ \frac{q-f(t)}{\gamma(t)} \right],
\end{equation}
where $\bs \lambda(t)=(\gamma(t), f(t))$ and $U_0[q]=U[q, \gamma(t)=1,f(t)=0]$. The parameter $\gamma$ represents dilation (expansion and
contraction) while $f$ represents translation. The class of systems
with the scale-invariant property covers a wide range of systems
such as those with square-well potentials and harmonic oscillator
potentials. These potentials keep overall shape
under $\bs \lambda(0) \to \bs \lambda (t)$. 

Let $\psi_n^0(q) = \bra q|n \ket$ be the $n$-th eigenfunction
of $H_0(\gamma=1, f=0)$. Then $\psi_n(q,\bs \lambda)
=\alpha(\gamma) \psi_n^0[(q-f)/\gamma]$ is an eigenfunction
of (\ref{eq:generalinv}) with the eigenvalue $E_n/\gamma^2$ and $\alpha(\gamma) =\gamma^{-1/2}$.
In fact, observe that
\begin{eqnarray*}
H(\bs \lambda)\psi_n(q,\bs \lambda) &=& -
\frac{\hbar^2}{2m} \partial_q^2  \alpha(\gamma)
\psi_n^0[(q-f)/\gamma] 
 +\frac{1}{\gamma^2} U_0((q-f)/\gamma) \alpha (\gamma) 
 \psi_n^0[(q-f)/\gamma] \\
 &=& \frac{\alpha(\gamma)}{\gamma^2}
 \left( -\frac{\hbar^2}{2m} \partial_\sigma^2
+U_0(\sigma)\right) \psi_n^0(\sigma) = \frac{E_n}{\gamma^2}
\psi_n(q,\bs \lambda),
\end{eqnarray*}
where $\sigma=(q-f)/\gamma$. The parameter $\alpha$ is fixed by the normalization
$1=\alpha(\gamma) ^2 \int dq \left[\psi_n^0((q-f)/\gamma) \right]^2
= \alpha(\gamma)^2 \gamma$ 
so that $\alpha(\gamma)=1/\sqrt{\gamma}$.

Now we rewrite $H_1$ by inserting the completeness relation of the coordinate basis $\{|q \ket\}$ as
\begin{eqnarray}
H_1(t)&=& i \hbar \dot{\bs \lambda}\cdot \sum_m
\left[|\nabla m\ket\bra m|-\bra m|\nabla m\ket |m \ket \bra m|
\right]
\nonumber\\
&=& i\hbar \dot{\bs\lambda} \cdot \sum_m \int dq 
\left[|q \ket \bra q| \nabla m\ket \bra m|
- \bra m|q \ket \bra  q|\nabla m \ket |m \ket \bra m| 
\right]\nonumber\\
&=& i\hbar \dot{\bs \lambda}\cdot
\sum_m \int dq |q \ket \nabla \psi_m(q, \bs \lambda) \bra m|
- i\hbar \dot{\bs \lambda} \cdot \sum_m \int dq
\bra m|q \ket \nabla \psi_m( q, \bs \lambda) |m \ket \bra m|,
\label{eq:h1aaa}
\end{eqnarray}
where $\nabla$ stands for $\nabla_{\bs \lambda}=(\partial/\partial \gamma, \partial/\partial f)$. Terms with nabla are simplified
by noting
$$
\nabla \psi_m(q, \bs \lambda) =
\left(\frac{\alpha'(\gamma)}{\alpha (\gamma)}\psi_m(q, \bs \lambda)
-\frac{q-f}{\gamma} \partial_q \psi_m(q,\bs \lambda),
-\partial_q \psi_m(q,\bs \lambda) 
\right).
$$
Then the bottom line of Eq.~(\ref{eq:h1aaa}) is written as
\begin{equation}
\left[\frac{\dot{\gamma}}{\gamma} (q-f) p + i \hbar \dot{\gamma}
\frac{\alpha'}{\alpha} + \dot{f} p\right]
- i \hbar \left[\frac{ \dot{\gamma}}{2 \gamma} + \dot{\gamma}
\frac{\alpha'}{\alpha}
\right].
\end{equation}
We finally obtain
\begin{equation}\label{eq:zzz}
H_1(\bs \lambda) = \frac{\dot{\gamma}}{2\gamma} [(q-f)p+p(q-f)] + \dot{f} p ,
\end{equation}
where use has been made of the canonical commutation 
relation $[q-f,p]=i \hbar$ to make $H_1(t)$ manifestly Hermitian. 
Note that $p$ is the generator of translation while $(pq+qp)$ is
the generator of dilation. In general, $H_1(t)$ of a Hamiltonian $H_0(t)$
with a symmetry is a linear combination of the generators of the symmetry.

Note that $H_1(t)$ thus obtained is 
non-local containing $pq+qp$, which is a challenge for a physical implementation. This can be solved, however, by applying a unitary transformation to make the Hamiltonian local\cite{ref:Ibanez,ref:adolfo} as we see in section 3 (c).

An example of a potential that satisfied the scaling requirement
is 
\begin{equation}
U[q, \gamma(t)]= \frac{A}{\gamma^2} \left( \frac{q}{\gamma}\right)^b
\end{equation}
where $b$ is a positive even integer and $A$ is a real positive
constant. This class contains a harmonic oscillator potential ($b=2$) and a square well potential ($b \to \infty$). 
Detailed analysis of the CD driving of a harmonic oscillator
will be made in the next section.

\section{Examples}

We introduce three examples to illuminate the formalism
developed in the previous section.

\subsection{Driving Spin}

Let us consider a spin $\boldsymbol{S} = (\sigma_x, \sigma_y, \sigma_z)/2$ in a
time-dependent magnetic field $\boldsymbol{B}_0=(B_x,B_y, B_z)$ with a Hamiltonian
\begin{equation}\label{eq:ham0}
H_0 (t) = \gamma \boldsymbol{B}_0(t) \cdot \boldsymbol{S},
\end{equation}
where $\sigma_k$ is the $k$th Pauli matrix and $\gamma$ is the gyromagnetic ratio. The two eigenvalues and corresponding normalized eigenvectors are 
\begin{eqnarray}
E_0(t) = - \gamma B_0(t)/2,&\quad& |E_0(t) \ket = 
\begin{pmatrix}
-\sin (\theta(t)/2)\\
e^{i \phi(t)} \cos(\theta(t)/2)
\end{pmatrix}
\\
E_1(t) =  \gamma B_0(t)/2,&\quad& |E_1(t) \ket = 
\begin{pmatrix}
\cos (\theta(t)/2)\\
e^{i \phi(t)} \sin(\theta(t)/2)
\end{pmatrix},
\end{eqnarray}
where $\theta(t)$ and $\phi(t)$ are polar coordinates of $\boldsymbol{B}_0(t)$ and $B_0(t)=|\boldsymbol{B}_0(t)|$.

It is interesting to see the dynamics of $H_0$ without $H_1$. Suppose
the Hamiltonian changes smoothly from $t=0$ to $t=T_0$. The evolution is adiabatic if $\gamma B_0 T_0/\hbar \gg 1$ and non-adiabatic otherwise.
Let us introduce the normalized time $t'$ by $t=T_0 t'$ so that
$0 \leq t' \leq 1$. Then the Schr\"odinger equation is rewritten as
\begin{equation}\label{scaled sch}
i \frac{d}{dt'} |\psi \ket 
= T \boldsymbol{b}_0(t')
\cdot \boldsymbol{S} |\psi \ket,
\end{equation}
where $T\equiv \gamma B_0 T_0/\hbar$ is the measure of adiabaticity and $\boldsymbol{b}_0 (t') = \boldsymbol{B}_0(t')/|\boldsymbol{B}_0(t')|$.
We write the normalized time $t'$ as $t$ from now on unless
it may cause confusion.
\begin{figure}\label{fig:nonad}
\begin{center}
\includegraphics[width= 10cm]{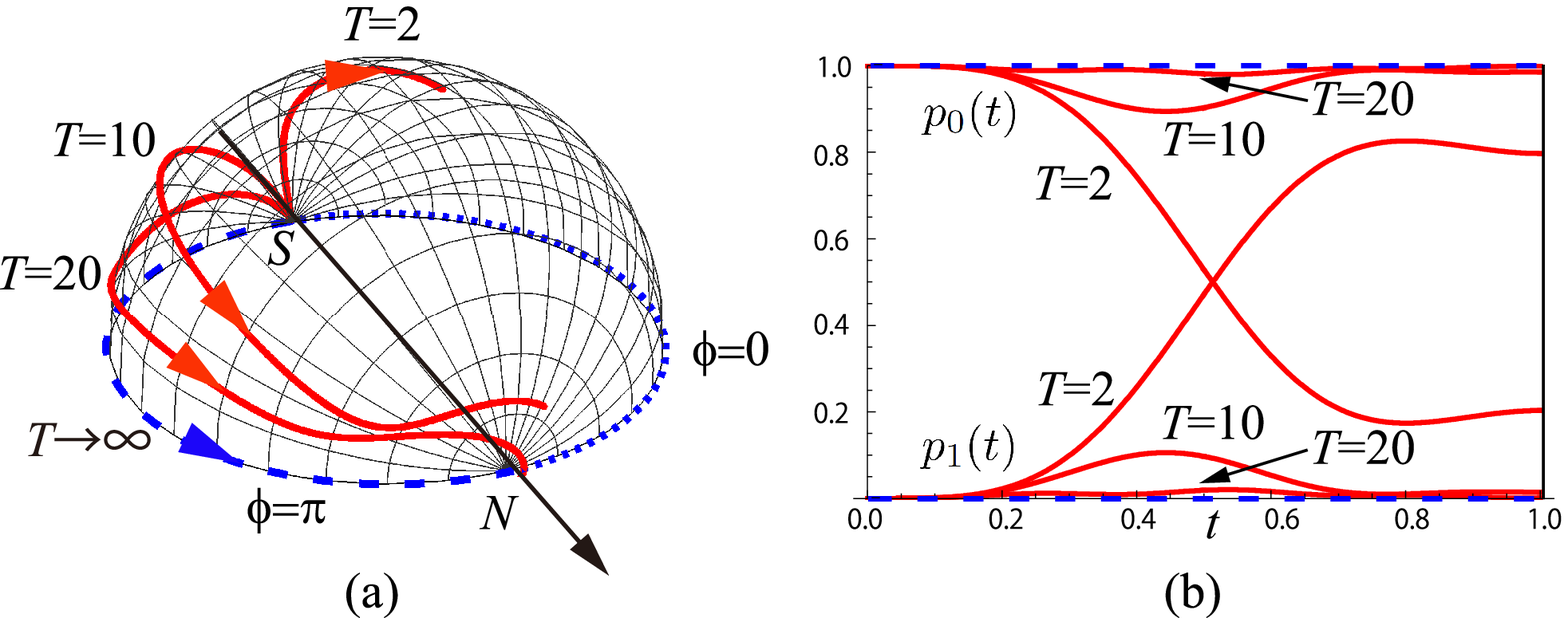}
\end{center}
\caption{(a) Trajectories of the Bloch vector
$\bra \psi(t)|\boldsymbol{\sigma}|\psi(t) \ket$ for $T=2,10, 20$
(sold red curves). The dashed blue curve $(\phi=\pi$) is the trajectory of adiabatic evolution while the dotted blue curve ($\phi=0$) shows $\boldsymbol{b}_0(t)$. $N$ and $S$ indicate the North and
the South Poles, respectively. The hemisphere is parameterized
as $0 \leq \theta \leq \pi$ and $0 \leq \phi \leq  \pi$.
(b) Populations $p_0(t)=|\bra E_0(t)|\psi(t) \ket |^2$ and $p_1(t)=|\bra E_1(t)|\psi(t) \ket |^2$ for the same choices of $T$. The dashed blue
lines show $p_0(t)=1$ and $p_1(t)=0$ for adiabatic evolution.
}
\end{figure}

Suppose the polar coordinates of $\boldsymbol{b}_0(t)$ change as
\begin{equation}\label{eq:polar}
\theta(t) = \pi \sin^2 \left( \frac{\pi t}{2}\right),\quad
\phi(t) =0,
\end{equation}
whose trajectory is the meridian $\phi=0$
from the North Pole ($t=0$) to the South Pole ($t=1$).
The Schr\"odinger equation (\ref{scaled sch}) is
solved with the initial condition $|\psi(0) \ket = |E_0 (0)\ket=(0,1)^t$.
The trajectory of the Bloch vector 
$\bra \psi(t)|\boldsymbol{\sigma}|\psi(t) \ket$ is depicted
in Fig.~1 (a) for $T=2, 10, 20$ and $T=\infty$ (adiabatic). The Bloch vector in the adiabatic limit
is always anti-parallel to $\bs{B}_0$. 
The populations $p_0(t)=|\bra E_0(t)|\psi(t) \ket |^2$ and $p_1(t)=|\bra E_1(t)|\psi(t) \ket |^2$ are shown
in Fig.~1 (b) for the same choices of $T$. The dashed blue lines
show the populations for the adiabatic evolution, $p_0(t)=1$ and
$p_1(t)=0$. Observe that how the Bloch vector trajectory
deviates from the adiabatic one and fails to reach the North
Pole at $t=1$ as $T$ is reduced. 
Note also that the populations $p_0(t)$ and $p_1(t)$ deviate
considerably from the adiabatic limit $p_0(t)=1, p_1(t)=0$ as $T$ is reduced. 
These observations justify the necessity of the CD term
for fast and precise control of states.

The CD Hamiltonian is found by using (\ref{eq:newh})
as 
\begin{eqnarray}
H_1&=& i  \partial_t \boldsymbol{B}_0 \sum_{j \neq k}
\frac{|E_j \ket \bra E_j|\boldsymbol{S}|E_k \ket \bra E_k|}
{E_k-E_j}\nonumber\\
&=& \frac{\partial_t \boldsymbol{B}_0}{2 B_0^2}\cdot
(B_{0z} \sigma_y-B_{0y} \sigma_z, 
B_{0x} \sigma_z-B_{0z} \sigma_x, 
B_{0y} \sigma_x-B_{0x} \sigma_y)\nonumber\\
&=& \frac{1}{2 B_0^2}
\big[ (B_{0y} \partial_t B_{0z}-B_{0z} \partial_t B_{0y} ) \sigma_x
+ (B_{0z} \partial_t B_{0x}-B_{0x} \partial_t B_{0z} ) \sigma_y\nonumber\\
& & \qquad +(B_{0x} \partial_t B_{0y}-B_{0y} \partial_t B_{0x}) \sigma_z\big] \nonumber\\
&=& \frac{1}{B_0^2} (\boldsymbol{B}_0 \times
\partial_t \boldsymbol{B}_0) \cdot \boldsymbol{S}
=  (\boldsymbol{b}_0 \times
\partial_t \boldsymbol{b}_0) \cdot \boldsymbol{S},
\end{eqnarray} 
where we suppressed explicit time dependence to simplify the expression.
It can be shown by using $\mathrm{tr}(\sigma_i \sigma_j) = 2 \delta_{ij}$ that $H_1$ is orthogonal
to $H_0$. Now the total Hamiltonian
$H_0 + H_1$ describes a spin in an effective magnetic field
\begin{equation}
\boldsymbol{B}= \boldsymbol{B}_0 + 
\frac{1}{\gamma} \boldsymbol{b}_0 \times \dot{\boldsymbol{b}}_0,
\end{equation} 
where the dot above $\boldsymbol{b}_0$ denotes the time derivative.
Let us scale time as before and introduce $T=\gamma B_0 T_0/\hbar$.
Then the Schr\"odinger equation is written as
\begin{equation}
i \hbar \frac{d}{d t}|\psi \ket =\left[
T \boldsymbol{b}_0
+ (\boldsymbol{b}_0 \times \dot{\boldsymbol{b}}_0
)\right] \cdot \boldsymbol{S} |\psi \ket.
\end{equation}
Observe that $H_1$ is independent of $T$ while $H_0$ is proportional
to $T$. The effect of the CD term becomes significant as $T$ is reduced.
For concreteness, let us take the polar coordinates (\ref{eq:polar})
and the same initial condition $|\psi(0) \ket= |E_0(0)\ket$ as before.
The CD term is found as
\begin{equation}\label{eq:spin1cd}
H_1(t)= \frac{\pi^2}{4} \sin(\pi t) \sigma_y.
\end{equation}
and the effective magnetic field is
\begin{equation}
\boldsymbol{B}(t) = \left( 
T \sin\left(\pi \sin^2\left( \frac{\pi t}{2}
\right)\right), \frac{\pi^2}{2} \sin (\pi t), T \cos\left(\pi \sin^2\left( \frac{\pi t}{2}
\right)\right)
\right).
\end{equation}
\begin{figure}
\begin{center}
\includegraphics[width=12cm]{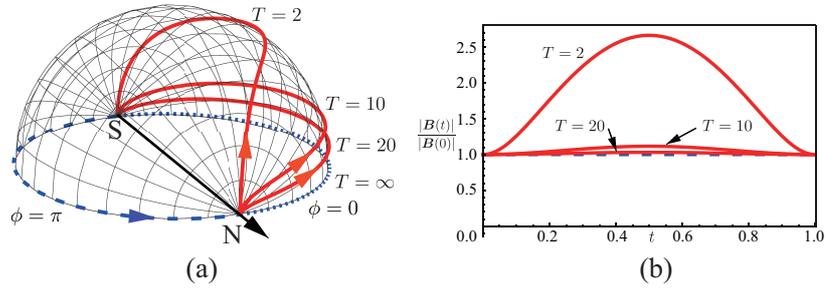}
\end{center}
\caption{(a) Trajectory of $\boldsymbol{b}(t)$ for $T=2,10,20$ and
$\infty$ (dashed curve). North and south poles of the Bloch sphere are denoted as $N$ and $S$, respectively. Dashed blue curve
($\phi=\pi$) is the trajectory of the Bloch vector for all $T$.  
(b) Amplitude $|\boldsymbol{B}(t)|/|\boldsymbol{B}(0)|$ 
for $T=2,10,20$ and
$\infty$ (dashed blue curve) as a function of $t$. Both trajectory and amplitude deviate from
those of $T=\infty$ (adiabatic limit) as $T$ is reduced.}\label{fig:1}
\end{figure} 
Figure \ref{fig:1}~(a)
shows the trajectory of $\boldsymbol{b}(t)=
\boldsymbol{B}(t)/|\boldsymbol{B}(t)|$ for
$T= 2, 10, 20$ and $\infty$ (adiabatic limit). Observe that the trajectory approaches
to that of $\boldsymbol{b}_0$ as $T$ becomes larger. 
Figure \ref{fig:1} (b) shows the normalized amplitude 
$|\boldsymbol{B}(t)|/|\boldsymbol{B}(0)|$ as a function of $t$ for $T=2,10,20$ and $\infty$.
The amplitude approaches the adiabatic
limit $1$ as $T$ increases. 

\begin{figure}
\begin{center}
\includegraphics[width=5cm]{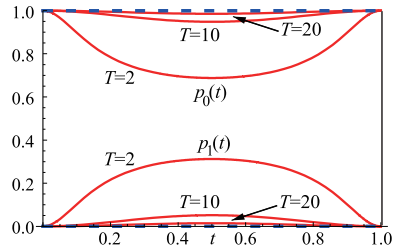}
\end{center}
\caption{Probabilities $p_0(t)=|\bra \mathcal{E}_0(t)|\psi(t) \ket|^2$ and
$p_1(t)=|\bra \mathcal{E}_1(t)|\psi(t) \ket|^2$ 
for $T=2,10,20$ and $\infty$ (dashed blue lines). Deviation of the
inner product from the adiabatic limit $p_0(t)=1$ and
$p_1(t)=0$ is significant for smaller $T$.}\label{fig:2}
\end{figure}

Let $|\mathcal{E}_0(t) \ket$ and $|\mathcal{E}_1 (t) \ket$ be the instantaneous 
eigenvectors  of $H(t)$ with eigenvalues $\mathcal{E}_0=-
\sqrt{4 T^2+\pi^4 \sin^2( \pi t)}/4$ and $\mathcal{E}_1=\sqrt{4 T^2+\pi^4 \sin^2( \pi t)}/4$,
respectively.
Then the transition between the two instantaneous 
eigenstates of $H(t)$ is characterized by the probabilities
\begin{equation}
p_0(t)= |\bra \mathcal{E}_0(t)|\psi(t) \ket|^2=
|\bra \mathcal{E}_0(t)|E_0(t) \ket|^2=
\frac{1}{2}
+ \frac{T}{\sqrt{4 T^2+\pi^4 \sin^2(\pi t)}}
\end{equation}
and
\begin{equation}
p_1(t)= |\bra \mathcal{E}_1(t)|\psi(t) \ket|^2=
|\bra \mathcal{E}_1(t)|E_0(t) \ket|^2=
\frac{1}{2}
- \frac{T}{\sqrt{4 T^2+\pi^4 \sin^2(\pi t)}}.
\end{equation}
Figure 3 shows 
$p_0(t)$ and $p_1(t)$ for $T=2,10,20$ and $\infty$
(dashed blue lines). The solution $|\psi(t)\ket$ of the Schr\"odinger equation remains
in $|E_0(t)\ket$ throughout the time evolution for any $T$, while
$|\psi (t)\ket$ deviates from $| \mathcal{E}_0(0) \ket$ 
except at $t=0$ and $t=1$ for a finite $T$.

\subsection{Driving Two Spins}

We will be sketchy in the next two examples to save space.
Readers are recommended to work out the details to
familiarize themselves with this subject.

Recently there has been a surge of applications of STA to
many-body systems, a part of which will be treated separately
in this issue. As a preliminary to such applications, we analyze
STA of a two-spin system, mainly based on \cite{ktaka}.

Let
\begin{equation}
H_0(t) = J_x(t) \sigma^1_x  \sigma^2_x + J_y(t) \sigma^1_y \sigma^2_y
+ h(t) (\sigma^1_x+\sigma^2_x)
\end{equation}
be a Hamiltonian of a two-spin system, where $\sigma^1_k =
\sigma_k \otimes I_2$ and $\sigma^2_k = I_2 \otimes \sigma_k$,
$I_2$ being the unit matrix of dimension 2.
$H_0$ is block-diagonalized in the basis
$\{ |00\ket, |11\ket, |01\ket, |10 \ket\}$ as
\begin{equation}
H_0=H_{01} \oplus H_{02}\ \ \mbox{where}
\ \ H_{01} = 2h \sigma_z + (J_x-J_y) \sigma_x,\ H_{02} = (J_x+J_y) \sigma_x.
\end{equation}
Here $\sigma_z|0\ket =|0 \ket$ and $\sigma_z|1 \ket=-|1 \ket$. 

Decomposition of $H_0$ into two single-spin Hamiltonians indicates that STA is reduced to that for single-spin systems. 
The CD Hamiltonian $H_{11}$ for the first block is
\begin{equation}
H_{11} = F  \sigma_y, \quad F=
\frac{h(\dot{J}_x-\dot{J}_y)-\dot{h}(J_x-J_y)}
{4h^2+(J_x-J_y)^2}
\end{equation}
while $H_{12}$ for the second block vanishes, which is obvious 
from the fact that the eigenvectors are independent of time.
$H_1=H_{11}$ in the original binary basis $\{|00 \ket, |01\ket, |10\ket, |11\ket\}$ is written as
\begin{equation}
H_1= \frac{1}{2} F (\sigma_x^1  \sigma_y^2+\sigma_y^1 \sigma_x^2).
\end{equation}
The resulting Hamiltonian is 
\begin{equation}\label{eq:satham}
H=H_0+H_1=\begin{pmatrix}
2h&0&0&-iF+J_x-J_y\\
0&0&J_x+J_y&0\\
0&J_x+J_y&0&0\\
iF+J_x-J_y&0&0&-2h
\end{pmatrix}
\end{equation}

The instantaneous eigenvalues and the corresponding eigenvectors of $H_0$ are 
\begin{equation}
\begin{array}{ll}
E_0= -\sqrt{4h^2+(J_x-J_y)^2}&|E_0\ket= \mathcal{N}_0
(2h +E_0, 0,0,J_x-J_y)^t\\
E_1 =-J_x-J_y& |E_1 \ket=(0,1,-1,0)^t/\sqrt{2}\\
E_2=J_x+J_y&|E_2 \ket=(0,1,1,0)^t/\sqrt{2}\\
E_3 =\sqrt{4h^2+(J_x-J_y)^2}&|E_3\ket= \mathcal{N}_3
(2h +E_3, 0,0,J_x-J_y)^t,
\end{array}
\end{equation} 
where $\mathcal{N}_{0,3}$ are normalization factors.
Let us concentrate on $|E_0 (t) \ket$ and consider adiabatic
time evolution during $-T \leq t \leq T$ such that $|E_0(-T) \ket =|0\ket|0\ket$ while
$|E_0(T) \ket=(|0 \ket |0 \ket - |1 \ket|1 \ket)/\sqrt{2}$
up to a phase.
Let us take control parameters
\begin{equation}
J_x= \tanh^2(v t),\ J_y=-\tanh(vt),\ h=\tanh(vt)-1.
\end{equation} 
Here $v$ and $T$ control adiabaticity.
For definiteness, we take $T=2$ and $v=5$
in the following, for which the boundary condition is satisfied
with a good precision. 
Figure 4 (a) depicts parameters $J_x(t), J_y(t)$ and $h(t)$ while (b) shows that spectra of $H_0(t)$ in the domain $-1 \leq t \leq 1$. The
spectra are essentially flat outside this domain. The blue solid
curve in Fig.~4~(c) shows the inner product $|\bra E_0(t)|\psi_0(t) \ket|^2$, where $|\psi_0(t) \ket$ is the solution of the 
Schr\"odinger equation with the Hamiltonian $H_0(t)$. Adiabaticity
is lost after $0<t$ and considerable amount of probability 
leaks to $|E_3(t) \ket$. There are no transitions
to $|E_1(t)\ket$
and $|E_2(t) \ket$ due to the block diagonal form of $H_0(t)$. 
\begin{figure}
\begin{center}
\includegraphics[width=13.5cm]{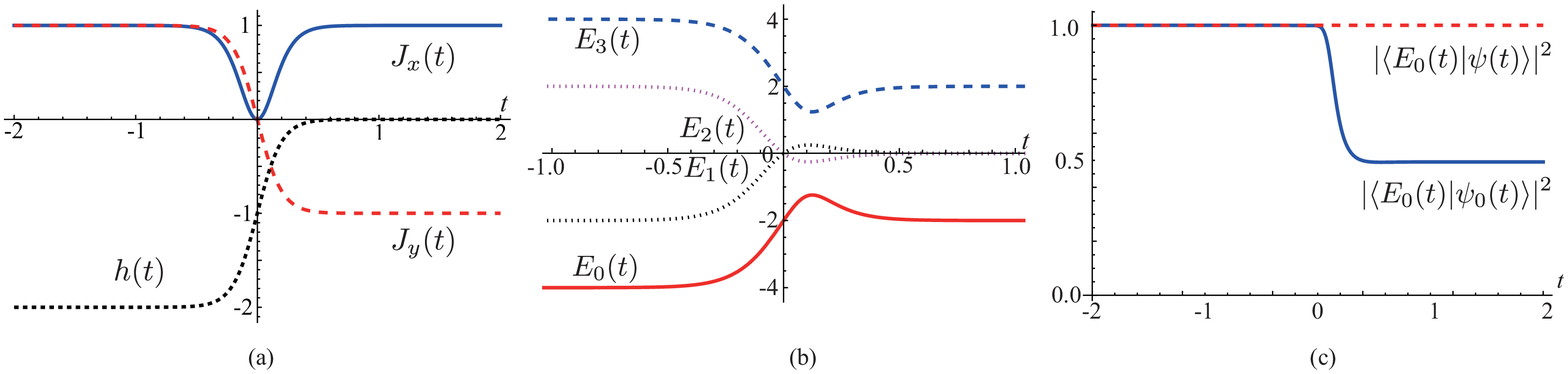}
\end{center}
\caption{(a) Parameters $J_x$ (solid blue curve), $J_y$ (dashed
red curve), and $h$ (dotted black curve) as functions of $t$. (bra $\{E_k(t)\}$ of $H_0(t)$ for $-1 \leq t \leq 1$. Spectra are constant in practice outside this domain. Solid red curve 
depicts $E_0(t)$
while dotted black, dotted magenta and dashed blue curves represent $E_1(t), E_2(t)$ and $E_3(t)$, respectively.
(c) Inner products $|\bra E_0(t)|\psi_0(t) \ket|^2$ and 
 $|\bra E_0(t)|\psi(t) \ket|^2$. There is a transition from $|E_0(t)\ket$ to $|E_3(t)\ket$ at $t \sim 0$ if $H_0(t)$ is employed.
 In contrast, the solution $|\psi(t)\ket$ of the Schr\"odinger equation with $H(t)$ reproduces the adiabatic solution $|E_0(t)\ket$ up to the phase.}
\end{figure}

We next solve
the Schr\"odinger equation with the Hamiltonian $H(t)$ with the initial condition $|\psi(-2)\ket=|E_0(-2)\ket=|0 \ket|0 \ket$, i.e., the ground state of $H_0(-2)$. The difference between the eigenvectors of $H_0(t)$ and
$H(t)$ is negligible at $t=-2$. The solution $|\psi(t)\ket$ 
thus obtained remains
in $|E_0(t) \ket$ so that $|\bra E_0(t)|\psi(t) \ket|^2=1$ for all $t$
as indicated by the red dashed curve in Fig.~4~(c). Moreover it is easily verified that
$|\psi(2)\ket = (|0 \ket|0 \ket - |1 \ket|1 \ket)/\sqrt{2}$.

\subsection{Harmonic Oscillator}

Examples so far have Hamiltonians described by  finite-dimensional matrices.
We next consider a harmonic oscillator, whose angular
frequency $\omega$ changes as a function of time.

Let us consider a harmonic oscillator with a Hamiltonian
\begin{equation}
H_0= \frac{p^2}{2m} + \frac{1}{2} m \omega(t)^2 q^2,
\end{equation}
where $\omega(t)^2= \omega(0)^2/\gamma(t)^4$.
Following the prescription outlined in section 2, the CD
Hamiltonian is derived as\footnote{This result was obtained first in
\cite{ref:mugax} without invoking the scale-invariance.} 
\begin{equation}\label{eq:h1harm}
H_1(t) = - \frac{\dot{\omega}}{4 \omega}(pq+qp).
\end{equation} 

Although $H_1(t)$ is Hermitian, its physical realization is challenging due to its nonlocality and we need more gadget to make it experimentally feasible.
For this purpose, we introduce a unitary transformation, with which $pq+qp$ can be eliminated. 
Let
\begin{equation}
U=\exp\left(\frac{i m \dot{\omega}}{4 \hbar \omega}q^2 \right).
\end{equation}
be a time-dependent unitary operator and let $\psi' (t)=U(t)^{\dagger}\psi(t)$. The Schr\"odinger equation that $\psi'(t)$
satisfies is $i \hbar \partial_t \psi'(t) = H'(t) \psi'(t)$, where
$H'(t)= U^{\dagger} H(t) U(t) - i U(t)^{\dagger} \dot{U}(t)$.
By using
$$
U^{\dagger}(t) p^n U(t) =\left(p+\frac{m \dot{\omega}}{2 \omega}q\right)^n\ \mbox{and}\ U^{\dagger}(t) q^n U(t) = q^n,
$$
one easily finds
\begin{equation}
H'(t) = \frac{p^2}{2m} + \frac{1}{2}m \omega'^2 q^2,
\quad  \omega' = \left(\omega^2 - \frac{3\dot{\omega}^2}
{4 \omega^2} + \frac{\ddot{\omega}}{2 \omega}\right)^{1/2}.
\end{equation}
This result was also obtained in \cite{ref:taka_unitary}. 

To make the analysis more concrete, let us consider $\omega(t)$ given by
\begin{equation}\label{eq:freqcd}
\omega(t)= \omega_0 \left(1 + \frac{1}{3}\tanh^3\frac{t}{T}\right)
\quad (0 \leq t \leq 5 T).
\end{equation}
It interpolates between $ \omega_0$ at $t=0$ and $\sim (4/3)\omega_0$ at $t=5 T$, where $T$ controls adiabaticity.
The parameters in (\ref{eq:freqcd}) have been chosen so that
$\omega'(t)$ remains positive real for $T=0.3$ and $\dot{\omega}$
and $\ddot{\omega}$ (approximately) vanish at $t=0$ and $5T$. 
$\omega(t)$ and $\omega'(t)$ are plotted in 
Fig.~5~(a) for $T=0.3$.
\begin{figure}\label{fig:omegax}
\begin{center}
\includegraphics[width= 10cm]{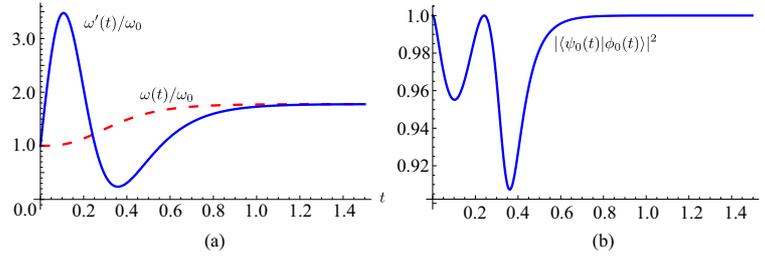}
\end{center}
\caption{(a) $\omega(t)$ (dashed red curve) and $\omega'(t)$
(solid blue curve) as functions of $t$ for $T=0.3$. (b) Overlap $|\bra \psi_0(t)|\phi_0(t) \ket|^2$ between the instantaneous ground state 
$|\phi_0(t) \ket$ of $H'(t)$ and the solution $|\psi_0(t)\ket$ of
the time-dependent Schr\"odinger equation with $H'(t)$ and
$|\psi_0(0)\ket=|\phi_0(0)\ket$, which reproduces Eq.~(\ref{eq:wv00}).
}
\end{figure}

It turns out to be convenient to introduce normalized time
$\tilde t=\omega_0 t$ and normalized coordinate $\tilde q
= q/l$, where $l= \sqrt{\hbar /m \omega_0}$. Then the 
original Schr\"odinger equation is written as
\begin{equation}\label{eq:scaledsch0}
i \frac{d \psi}{d\tilde t}=\left( -\frac{1}{2} \frac{d^2}{d\tilde t^2} 
+ \frac{1}{2} \left(2 + \tanh^3\frac{t}{T}\right)^2 \tilde q^2\right) \psi.
\end{equation}
The instantaneous ground state of the above Hamiltonian is
\begin{equation}\label{eq:wv00}
\psi_0(\tilde q, t) = C(t) e^{-\tilde q^2/2}\qquad 
C(t) = \left(\frac{m \omega(t)}{\pi \hbar}\right)^{1/4}. 
\end{equation}
The solution of the Schr\"odinger equation with $H'(t)$
is
$\psi_0(\tilde q,t)$.
Figure~5~(b) shows the overlap
$|\bra \psi_0(t)|\phi_0(t) \ket|^2$ for $T=0.3$, 
where $\phi_0(t)$ is the instantaneous ground state of 
$H'(t)$, in which 
$\omega(t)$ in Eq.~(\ref{eq:wv00}) is replaced by $\omega'(t)$.
Clearly $|\bra \psi_0(t)|\phi_0(t) \ket|^2=1$ whenever
$\omega(t) =\omega'(t)$.


\section{Another View of Counterdiabatic Driving}
 
We have introduced the CD driving in section~2
through the spectral decomposition of $H_0$. Here, we derive
the CD Hamiltonian $H_1$ from a different viewpoint
based on \cite{ref:sp,ref:spr}. This formalism also provides with
approximate variational $H_1$ even when it is impossible
to obtain an exact $H_1$.

Let $H_0$ be a Hamiltonian with time-dependent parameters
$\bs \lambda(t)$. $H_0$ is diagonalized by employing the
instantaneous eigenvector basis $\{|n\ket\}$.
For definiteness, we assume there is only one parameter
$\lambda(t)$ and replace $\bs \lambda$ by $\lambda$. 
Generalization to multiple parameters is obvious.

Let $U(\lambda(t))$ be the unitary transformation associated with the basis change so that $\tilde H_0=U^{\dagger}H_0 U$ is diagonal. 
In this ``moving frame'', a state transforms as $|\psi \ket
\to |\tilde \psi \ket = U^{\dagger}|\psi \ket$. 
The Schr\"odinger equation in the moving frame is 
\begin{equation}
i \hbar \frac{d}{dt}|\tilde \psi \ket = (\tilde H_0 - 
\dot{\lambda} \tilde \cA_{\lambda})|\tilde \psi \ket,
\end{equation} 
where $\tilde \cA_{\lambda} = -i \hbar (\partial_{\lambda}U^{\dagger}
)U = i \hbar U^{\dagger} \partial_{\lambda} U$ is called
the adiabatic gauge potential with respect to $\lambda$.
$\tilde H_{m0} =\tilde H_0 - \dot{\lambda} \tilde \cA_{\lambda}$
is the Hamiltonian in the moving frame.
Note that $\tilde H_{m0}$ is not diagonal in general due to the second term. This is why transitions among eigenstates of $\tilde H_0$ take place. 

Now what we have to do to prevent transitions among eigenstates
of $\tilde H_0$ should be clear. Let us introduce the CD term
in the moving frame by
\begin{equation}
\tilde H_1 = \dot{\lambda} \tilde \cA_{\lambda}.
\end{equation}
Then the total Hamiltonian 
\begin{equation}
\tilde H_m = \tilde H_0 -\dot{\lambda}\tilde \cA_{\lambda} +  
\tilde H_1 = \tilde H_0
\end{equation}
is diagonal and time-evolution due to
$\tilde H_m$ is transitionless. 
$\tilde H_1$ in the lab frame is
\begin{equation}
H_1 =U \tilde H_1 U^{\dagger}
= i \hbar  \dot{\lambda} (\partial_{\lambda} U) U^{\dagger}.
\end{equation}

\subsection{Examples}

Let us re-derive the CD Hamiltonian of some of the examples
introduced previously. 

We first consider a single spin in
a magnetic field $\bs B_0(\theta, \phi)$ with polar coordinates
$\theta(t)= \pi \sin^2( \pi t/2), \phi(t)=0$ $(0 \leq t \leq 1)$.
The Hamiltonian (\ref{eq:ham0}) is diagonalized by
\begin{equation}\label{eq:unitaryx}
U= \left(
\begin{array}{cc}
-\sin \left({\theta}/{2}\right)&   \cos \left({\theta}/{2}\right) \\
 e^{i \phi } \cos
   \left({\theta}/{2}\right)& e^{i \phi } \sin \left({\theta}/{2}\right) 
\end{array}
\right)
\end{equation}
as 
\begin{equation}
\tilde H_0 = U^{\dagger} H_0 U = -B_0 \sigma_z.
\end{equation}
The CD Hamiltonian is evaluated as
\begin{equation}
\tilde H_1 = \dot{\theta(t)} i \hbar U^{\dagger} \partial_{\theta}
U = -\frac{\hbar}{4} \pi ^2 \sin (\pi  t) \sigma_y.
\end{equation}
$\tilde H_1$ is transformed back to the lab frame as
\begin{equation}
H_1 = U \tilde H_1 U^{\dagger} =\frac{\hbar}{4} \pi ^2 \sin (\pi  t) \sigma_y,
\end{equation}
in agreement with Eq.~(\ref{eq:spin1cd}).

In section 3, a harmonic oscillator with time dependent trap 
frequency was considered. Here we analyze a harmonic potential
with a moving center, whose Hamiltonian is 
\begin{equation}
H_0 = \frac{\hat{p}^2}{2m} + \frac{1}{2} m \omega^2 (\hat{q}-q_0)^2
\end{equation}
where $q_0$ is a function of time. We use $\hat{q}$ and $\hat{p}$ to indicate they are operators. We need a unitary transformation that maps
$(\hat{q}-q_0)^2$ to $\hat{q}^2$ to diagonalize the Hamiltonian, namely $U^{\dagger} (\hat{q}-q_0)^n U= \hat{q}^n$.
We find 
$U(q_0) = e^{-i \hat{p} q_0/\hbar}$ does diagonalize $H_0$ as
\begin{equation}
\tilde H_0 = U^{\dagger}(q_0) H_0 U(q_0) = 
\frac{\hat{p}^2}{2m} + \frac{1}{2} m \omega^2 \hat{q}^2
=(a^{\dagger} a + 1/2) \hbar \omega 
\end{equation}
where $a^{\dagger}$ and $a$ are ordinary creation and annihilation 
operators.

Now $\tilde H_1$ is evaluated as
\begin{equation}
\tilde H_1 = \dot{q}_0 i \hbar U(q_0)^{\dagger}\partial_{q_0} U(q_0)
=  \dot{q}_0 \hat{p}
\end{equation} 
in the moving frame while 
\begin{equation}
H_1 = \dot{q}_0 i \hbar U(q_0) \hat{p} U(q_0)^{\dagger}= \dot{q}_0 \hat{p}
\end{equation} 
in the lab frame, in agreement with Eq.~(\ref{eq:zzz}).

The total Hamiltonian in the lab frame is
\begin{equation}
H=\frac{\hat p^2}{2m} + \frac{1}{2} m \omega^2 ( \hat{q}
-q_0)^2 + \dot{q}_0
\hat{p}=\frac{(p+m \dot{q}_0)^2}{2m} + \frac{1}{2} m \omega^2 ( \hat{q}-q_0)^2-\frac{m \dot{q}_0^2}{2}.
\end{equation}
We introduce a gauge transformation $\hat{p} \to \hat{p}+\partial_q
f$ and $H \to H+\partial_t f$ with $f=-m \dot{q}_0 \hat{q}/\hbar$
so that 
\begin{equation}
H \sim \frac{\hat p^2}{2m} + \frac{1}{2} m \omega^2 ( \hat{q}
-q_0)^2 - m \ddot{q}_0 \hat{q},
\end{equation}
which is physically feasible. Here $\sim$ denotes the gauge-equivalence of both sides.

\subsection{More on Adiabatic Gauge Potentials}

Let us look at the spin-1/2 example in section 3~(a)
again. The unitary matrix (\ref{eq:unitaryx}) is written as
$U=(|E_0(t) \ket, |E_1(t) \ket)$ and $\tilde \cA_{\theta}$ as
\begin{equation}\label{eq:matrixa}
\tilde \cA_{\theta} =i \hbar  U^{\dagger}\partial_{\theta}U
= i \hbar \begin{pmatrix}
\bra E_0(t)|\partial_{\theta}|E_0(t)\ket & \bra E_0(t)|\partial_{\theta}|E_1(t)\ket
\\
\bra E_1(t)|\partial_{\theta}|E_0(t)\ket & \bra E_1(t)|\partial_{\theta}|E_1(t)\ket
\end{pmatrix}
\end{equation}
In this sense, Eq.~(\ref{eq:matrixa}) is regarded as a matrix expression of the
operator $\cA_{\theta} = i \hbar \partial_{\theta}$. 

The diagonal component of $\cA_{\lambda}$ is nothing but the Berry connection
\begin{equation}
A_{\lambda}^{(n)} = 
\bra n(\lambda)| \cA_{\lambda} |n(\lambda)\ket = i \hbar
\bra n(\lambda)|\partial_{\lambda} |n(\lambda)\ket.
\end{equation} 
Since $\{|n(\lambda) \ket\}$ is the set of eigenvectors of $H_0(\lambda)$, we have
\begin{equation}
\bra m(\lambda)|H_0(\lambda)|n(\lambda)\ket=0 \quad (m \neq n).
\end{equation}
By differentiating this with respect to $\lambda$, we obtain
$$
0 = (\partial_{\lambda}  \bra m|)H_0|n \ket 
+ \bra m|(\partial_{\lambda} H_0) |n \ket + \bra m|H_0
(\partial_{\lambda} |n \ket)=-\frac{i}{\hbar} (E_m-E_n)\bra m|\cA_{\lambda}
 |n \ket + \bra m|\partial_{\lambda} H_0 |n \ket,
$$
from which we find the matrix elements of $\partial_{\lambda}
H_0$ in terms of those of $\hat \cA_{\lambda}$ as
\begin{equation}
\bra m|\partial_{\lambda} H_0 |n \ket =
\frac{1}{i \hbar} ( E_n -E_m) \bra m| \cA_{\lambda}|n \ket 
 \quad (m \neq n).
\end{equation}
For the diagonal entries, we find $ \bra n|\partial_{\lambda}H_0 |n\ket = \partial_{\lambda}( \bra n|H_0|n \ket)= \partial_{\lambda} E_n$. By collecting these matrix elements,
we prove the operator identity
\begin{equation}
i \hbar \partial_{\lambda}  H_0 = [\cA_{\lambda} , H_0]- i \hbar 
F_\mathrm{ad}\quad \mbox{with}\quad
F_\mathrm{ad} = - \sum_n (\partial_{\lambda} E_n)|n\ket \bra n|.
\end{equation}
Evaluation of $F_\mathrm{ad}$ requires the spectral decomposition 
of $H_0(t)$, which we want to avoid as much as possible.
Luckily, we may take advantage of the identity $[H_0, F_\mathrm{ad}]=0$ to derive
\begin{equation}\label{eq:identity}
[H_0, i \hbar \partial_{\lambda} H_0-[ \cA_{\lambda} , H_0]]=0,
\end{equation}
By multiplying $\dot{\lambda}$, the above equality is put
in a form
\begin{equation}\label{eq:identity}
[H_0, i \hbar \partial_{t} H_0-[H_1 , H_0]]=0,
\end{equation}
which can be used to find $H_1$ without employing the spectral decomposition of $H_0$. Equation~(\ref{eq:identity}) can be also
used to derive an approximate $H_1$ by variational principle. Namely, introduce first an Ansatz of
$H_1$ with parameters and then minimize the operator norm
of the left-hand side of Eq.~(\ref{eq:identity}) with respect to the parameters to find an approximate $H_1$.  
See \cite{ref:sp,ref:spr} for details.

\section{Couterdiabatic Driving and Dynamical Invariant}

The CD formalism has a close relationship with another formalism of STA based
on the dynamical invariant (DI for short)\cite{equiv}. Let us give a brief introduction to the latter to begin with. 

\subsection{Dynamical Invariant}

Let $\cH(t)$ be a time-dependent Hamiltonian and $I(t)$ be a time-dependent Hermitian operator
both acting on $\mathbb{C}^d$. We require they satisfy
\begin{equation}\label{eq:dinv}
i \hbar \frac{\partial I(t)}{\partial t}= [\cH(t), I(t)].
\end{equation}
The operator $I(t)$ is called the dynamical invariant or the 
Lewis-Riesenfeld invariant \cite{lewis}. 
Let $|\chi(t) \ket$ be a solution of the Schr\"odinger equation
\begin{equation}\label{eq:schdi}
i \hbar \frac{d}{dt}|\chi(t) \ket = \cH(t)|\chi(t) \ket.
\end{equation}
It can be shown by using (\ref{eq:dinv}) that
\begin{equation}
\frac{d}{dt} \bra \chi(t)|I(t)|\chi (t) \ket = 0.
\end{equation}

Let $\{\lambda_n\}$ be the set of eigenvalues of $I(t)$ and
$\{|\phi_n(t) \ket\}$ be the corresponding set of normalized eigenvectors; $I(t)|\phi_n(t) \ket= \lambda_n|\phi_n(t) \ket$.
Note that the $t$-dependence of $\lambda_n$ is dropped since
it can be shown that $d \lambda_n/dt=0$. As a result, $I(t)$
has the following spectral decomposition
\begin{equation}\label{eq:iii}
I(t)= \sum_n \lambda_n |\phi_n(t) \ket \bra \phi_n(t)|, \quad \lambda_n \in \mathbb{R}.
\end{equation}

Take $|\phi_n(0) \ket$ and consider a solution $|\chi_n(t)\ket$
of Eq.~(\ref{eq:schdi}) 
such that $|\chi_n (0) \ket = |\phi_n(0)\ket$. Note that
the index $n$ in $|\chi_n(t) \ket$ does not indicate it is the
$n$th instantaneous eigenvector of $\cH(t)$ but rather
its inital state is $|\phi_n(0)\ket$. 
It is shown that $|\chi_n (t) \ket$ is written as 
\begin{equation}\label{eq:solnx}
|\chi_n (t) \ket = e^{i \alpha_n(t)} |\phi_n(t) \ket
\quad \mbox{with}\quad \alpha_n(t) =\frac{1}{\hbar} \int_0^t \bra \phi_n(s)|(i \hbar \partial_s -\cH(s))
|\phi_n(s) \ket ds.
\end{equation}
This expression should be compared with Eq.~(\ref{eq:adia}),
which is also written as
$$
|\psi_n(t) \ket = e^{i \xi_n(t)}  |n(t) \ket,\quad
\xi_n(t) =\frac{1}{\hbar} \int_0^t  \bra n(s)| ( i \hbar \partial_s-H(s))| n(s) \ket ds,
$$
where $|n(t)\ket$ is an eigenvector of $H_0(t)$. Note that
this is regarded as a special case of Eq.~(\ref{eq:solnx}).
Here $H_0(s)$ in Eq.~(\ref{eq:adia}) can be replaced by $H(s)=
H_0(s)+H_1(s)$ since $\bra n(s)|H_1(s)|n(s) \ket=0$.

Let $|\chi(t) \ket$ be an arbitrary solution of (\ref{eq:schdi}). Since $\{|\phi_n(0)\ket\}$ is a complete set, $|\chi(0)\ket$ can be expanded as
$$
|\chi(0)\ket = \sum_n c_n |\phi_n(0)\ket, \qquad 
c_k = \bra \phi_n(0)|\chi(0)\ket.
$$ 
By linearity, the solution at arbitrary $t > 0$ is 
\begin{equation}\label{eq:arbsol}
|\chi (t) \ket = \sum_n c_n e^{i \alpha_n(t)}|\phi_n(t) \ket.
\end{equation}
Observe that the set $\{c_n\}$ is independent of time, which means
the time-evolution of $|\chi(t)\ket$ is transitionless
in terms of the eigenvectors of $I(t)$. Since $I(t)$
and $\cH(t)$ do not commute with each other in general, 
$|\chi(t) \ket$ undergoes transitions among instantaneous 
eigenvectors of $\cH(t)$ and hence the time-evolution is nonadiabatic.
%

Let us write $|\chi(t) \ket = \cU(t)|\chi(0)\ket$, where
$\cU(t) = \mathcal{T} e^{-i\frac{1}{\hbar} \int_0^t \cH(s) ds}$ is the time-evolution
operator associated with $\cH(t)$ and $\mathcal{T}$ is the time-ordering operator. Since $\cU(t)$ maps $|\chi_n(0)\ket$ to 
$|\chi_n(t) \ket$, it is expressed as
\begin{equation}\label{eq:U}
\cU(t)=\sum_n |\chi_n (t) \ket \bra \chi_n(0)|
= \sum_n e^{i \alpha_n(t)} |\phi_n(t) \ket \bra \phi_n(0)|. 
\end{equation}
This confirms again that the evolution of 
$|\phi_n(t) \ket$ is transitionless since $\cU(t) |\phi_n(0)\ket = e^{i \alpha_n(t)}|\phi_n(t)\ket$ at any $t>0$. 

\subsection{Relation between Couterdiabatic Driving and Dynamical Invariant}

By inserting (\ref{eq:U}) into the Schr\"odigner 
equation $i \partial_t \cU(t) = \cH(t) \cU(t)$, we can inversely define a Hamiltonian
\begin{eqnarray}
\cH(t)& =& i \hbar (\partial_t \cU)\cU^{-1} 
= -\hbar \sum_n \dot{\alpha}_n(t)|\phi_n(t) \ket \bra \phi_n(t)|
+ i \hbar
 \sum_n |\partial_t \phi_n(t) \ket \bra \phi_n(t)|\nonumber\\
&=& F(t) +  i \hbar \sum_n |\partial_t \phi_n(t)\ket \bra \phi_n(t)|,\label{eq:diham}
\end{eqnarray} 
where 
$F(t)=-\hbar \sum_n \dot{\alpha}_n(t) |\phi_n(t) \ket \bra \phi_n(t)|$.
It is straightforward to verify (\ref{eq:diham}) with (\ref{eq:iii})
satisfies (\ref{eq:dinv}).
Note that $\cH(t)$ has some degrees of freedom originating from
the choice of $\alpha_n(t)$. If we demand $[\cH(0), I(0)]=
[\cH(T), I(T)] =0$, $\cH(t)$ and $I(t)$ have simultaneous eigenvectors
at $t=0$ and $T$ 
and state transfer is realized with no final excitations.
A convenient, although not necessary, choice is to set $I(0)=\cH(0)$. (\ref{eq:diham}) should be compared with (\ref{eq:hamcd}) written
in the form
\begin{equation}
H(t) = G(t) + i \hbar \sum_n |\partial_t n(t) \ket \bra n(t)|,
\quad G(t) =-\hbar \sum_n \dot{\xi}_n(t) |n(t) \ket \bra n(t)|.
\end{equation}
The Hamiltonian (\ref{eq:diham}) is for finite dimensions, whereas DI was originally introduced for infinite dimensions \cite{lewis}. 

Now the relation between the CD formalism and
the DI formalism should be clear. 
If we define $\cH_0(t)$ as the diagonal part of $\cH(t)$, we obtain
\begin{equation}
\cH_0(t) = \hbar \sum_n  [i \bra \phi_n(t)|\partial_t \phi_n(t)\ket - \dot{\alpha}_n(t)]|\phi_n(t) \ket \bra \phi_n(t)|.
\end{equation}
If we identify $\alpha_n(t)$ with $\xi_n(t)$ and $|\phi_n(t) \ket$
with $|n(t)\ket $, we have $G(t) = F(t)$ and $H(t)$ in 
(\ref{eq:hamcd}) and $\cH(t)$ in (\ref{eq:diham}) are identified. 
Moreover $H_0(t)$ is identified with
$I(t)$ if $E_n(t)$ is independent of time (i.e., isospectral change) and put $\lambda_n=E_n(0)$. In this way, the dynamical mode $|\phi_n(t) \ket$ associated with $I(t)$ is identified with the adiabatic mode $|n(t) \ket$ of $H_0(t)$.

It should be emphasized again that the two formalisms need not to be identical. There is a family of Hamiltonians interpolating between $\cH(0)$ and $\cH(T)$. The two formalisms are identified only with the particular choices $E_n=\lambda_n$,
$\alpha_n(t)=\xi_n(t)$ and $|\phi_n(t) \ket=|n(t)\ket$.

\section{Applications and Demonstrations of CD Formalism}

As noted in the beginning, the objective of this article
is to give the reader enough background to access research papers on CD formalism, including some of the articles in
this issue, rather than exhausting all the subjects related
to the formalism. Nonetheless, we will have a brief look at 
several applications and demonstrations of the CD formalism
so that this article also serves as a mini-review and 
a guide to further reading.
Since the CD formalism has been applied in wide areas in
physics, chemistry and control theory and the number of pages assigned to this article is limited, we must choose subjects to be introduced in this section. We apologize in advance to authors whose works are not mentioned here.

\subsection{CD Driving in Open Quantum Systems}
 
The CD formalism is not restricted within closed systems whose
time-evolution is described by unitary operators. 
Note that the gap between the neighboring energy eigenvalues
does not define the time-scale of adiabaticity 
due to the interaction between the system and the environment

It is possible to write the Gorini-Kossakowski-Lindblad-Sudarshan
(GKLS) equation $\dot{\rho} = \mathcal{L}[\rho]$ as an ordinary
matrix equation by introducing the orthonormal basis $\{ e_k\}_{1 \leq k \leq D^2}$ of matrices acting on $\mathbb{C}^D$\cite{ref:Sarandy,openq1}.
Then a density matrix $\rho = \sum_k \rho_k e_k$ is vectorized
as $|\rho\ket\ket = \{\rho_k\}_{1 \leq k \leq D^2}$ and
the superoperator $\mathcal{L}$ is represented as 
a $D^2 \times D^2$ supermatrix $L(t)_{jk}
= \mathrm{Tr}(e_j^{\dagger} \mathcal{L}[e_k])$, with which
the GKLS equation takes the form $|\dot{\rho}\ket\ket= L(t)
|\rho\ket\ket$.
$L(t)$ can be put in the Jordan canonical form by a similarity
transformation $C(t)$ as $L_J(t) = C^{-1}(t) L(t) C(t) = 
\mathrm{diag}(J_1(t), \ldots, J_N(t))$. 
The time-evolution may be defined as adiabatic if
there are no transitions among the Jordan blocks $\{J_s\}$. The similarity transformation $C(t)$ introduces
inter-block couplings through $\dot{C}(t)$, which may
be cancelled by the CD term that suppresses
the nonadiabatic transitions.

CD driving in open systems has been also analyzed in
\cite{openq2}, where a target trajectory of the evolution
of the system is given first and then required CD Hamiltonian
and dissipators are determined. The result is
interpreted as a driven system in the presence of balanced gain and loss which can be attributed to $\mathcal{PT}$-symmetric
quantum mechanics. It can be also implemented via a 
non-Markovian evolution in which a generalized GKLS equation describes the dynamics. 

Separating heat and work of a thermal process of an open
quantum system is ambiguous. Let $\rho(t)$ be a density matrix of an open system,
which is regarded as a trajectory in the state space. 
Associated with a change $d U$ of the internal energy, \cite{sahar} defined the heat change $d'\mathbb{Q}$ as an entropy-related
contribution while the work change $d'\mathbb{W}$ as a part
causing no entropy change, where $d'$ indicates it is not exact. 
They employed the ``trajectory-based STA'' (TB-STA)
\cite{openq2}
to describe the trajectory of $\rho(t)$ with a GKLS-like
equation, whose ``Hamiltonian'' takes the CD form
and the ``Lindbladian'' introduces jumps between the instantaneous 
eigenbasis. They have shown that the dissipative and coherent
parts of this equation contributed to heat and work, respectively.

\subsection{Quantum Speed Limit and Cost of CD Driving}

It seems at first sight that STA, including CD driving, can 
accelerate quantum control indefinitely so that an initial state 
can be shuttled to a final state in an arbitrarily short time.
However it has been know that there is so-called 
the quantum speed limit (QSL for short) $\tau_\mathrm{QSL}$
that defines the lower bound of the time required for a quantum
process. See \cite{qsl review} for a review on QSL. 
Roughly speaking QSL is given by the distance between the
initial and the final states divided by energy fluctuation or by expectation value of energy.
The trade-off between speed and energetic cost in the context of
STA was discussed initially in \cite{ref:Santos,qsl sta}.
They related the cost of CD driving\cite {cost} and QSL
and elucidated trade-off between the two.
The cost of CD driving is also discussed in 
\cite{frictionless}

An inequality between 
the nonequilibrium work fluctuations and the operation time was
derived in \cite{funok}, which 
shows speed-up by CD driving requires large fluctuation
in work, which is identified as the thermodynamic cost.
This theory is experimentally demonstrated with a
superconducting Xmon qubit \cite{funoexp}.

\cite{funox} considered the GKLS equation to derive QSL of open
quantum systems. The trade-off relation between the operation time and
the physical quantities such as
the energy fluctuation of the system Hamiltonian and the CD Hamiltonian and the entropy production was discovered. 

QSL is a general concept not restricted within the CD driving.
However it was conjectured that QSL for all fast-forward protocols are
bounded by that for CD driving \cite{bukov}. 
They validated this conjecture with a three-level system,
nonintegrable spin chains and the Sachdev-Ye-Kitaev model.

Measurement of QSLs in an ultracold quantum gas confined in a time-dependent harmonic trap was considered \cite{qsl measure}. 
It was shown
that QSL can be probed whenever the dynamics was self-similar by measuring the cloud size of the trapped gas as a function of
time. The Bures length and energy fluctuations that determine
QSLs are measured by this.
 
The CD control of a spin-boson model was analyzed 
to derive an upper bound on the performance and it was shown that unit fidelity can be reached by a time-dependent control of the interaction called the exact STA protocol \cite{funo1}.

\subsection{STIRAP with Cold Atoms}

STIRAP (STImulated Raman Adiabatic Passage)
is a protocol for population transfer between two quantum states
by employing typically two coherent pulses and the third quantum state that is not occupied during the process. The three quantum
states form the so-called $\Lambda$-system. 
This protocol is known to be robust against control errors
but takes long time since the control must be adiabatic.

STA was proposed to speed up STIRAP theoretically
\cite{LCtheory} and
demonstrated experimentally using cold atoms \cite{atomstirap},
where CD driving with a unitary transformation was employed
to propose STIRSAP (STImulated
Raman Shortcut-to-Adiabatic Passage),
which can be implemented by modulating the shape of Raman
pulses. Demonstration has been done by employing $|1 \ket = 
|F=1, m_F=0\ket$ and $|2 \ket = |F=2, m_F=0 \ket$ ground states
with additional $|3 \ket = |5^2\mathrm{P}_{3/2}\ket$ state
of $^{87}$Rb atom. With the operation time $0.4$~ms, 
STIRSAP attained the transfer efficiency of $100\%$ while 
STIRAP attained only $36\%$.

\subsection{Superconducting Qubits}

A quantum computer with superconducting qubits is one of the
most promising platforms of a scalable quantum computer.
Naturally there are many proposals and demonstrations of 
STA employing superconducting qubits. The number of qubits
in the NISQ is not large enough to incorporate quantum error correcting codes and hence it is essential that the gate operation time is shortened by STA to fight against decoherence. Among many physical realizations, the transmon qubit is advantageous over the other proposals due to its robustness against charge noise.

Speedup of the adiabatic population transfer in a three-level superconducting transmon circuit has been experimentally demonstrated \cite{sc1}.
The STIRAP protocol that realizes fast and robust population transfer from the ground state to the second excited state
has been accelerated with an additional two-photon microwave 
pulse implementing the CD Hamiltonian.

Speed and fidelity are further improved by combining
STIRSAP with optimal control theory (STIRSAP-Opt). \cite{sc1.1}
demonstrated this with four levels of a transmon qudit.

The CD driving is applied to an open 
superconducting circuit QED system with multiple lossy modes
coupled to a transmon and demonstrated the adiabatic evolution time of a single lossy mode was reduced from 800~ns to 100~ns 
\cite{sc2}. It was also demonstrated that an optimal control protocol realized fast and qubit-unconditional equilibrium of multiple lossy modes.

\subsection{Trapped-Ion Displacement}

It was shown in section 4 that a particle in a harmonic potential
can be transported with high-speed by CD driving.
It was demonstrated experimentally that a trapped-ion displacement in the phase space can be accelerated by CD driving. 
Suppose one wants to transport an ion in a harmonic trap
$V= m \omega^2 \hat{x}^2/2$ by 
adding a term $f(t) \hat{x}$, which shifts the 
potential by $q(t) =  -f(t)/m \omega^2$. 
A unitary transformation $U(t) = e^{i q(t) \hat{p}(t)}$ is introduced
to transform to the comoving frame to eliminate $f(t) \hat{x}$. This will introduce an additional
term $\dot{f}(t)\hat{p} /m \omega^2$ in the Hamiltonian,
which causes diabatic transitions. This term is eliminated
by adding the CD Hamiltonian  $H_\mathrm{CD}=-\dot{f}(t)\hat{p} /m \omega^2$. This is experimentally demonstrated by using a
trapped ${}^{177}\mathrm{Yb}^+$ ion \cite{iontrap}. 
This protocol is equivalent
to the result of the quantum brachistochrone solution \cite{takaqb}.

\subsection{Creation of Topological Excitations in BEC}

Vortices in the Bose-Einstein condensate (BEC) of alkali metal atoms with hyperfine spin degrees of freedom can be created by imprinting the Berry phase on a uniform condensate \cite{becv1,becv2,becv3}. This proposal was
subsequently demonstrated \cite{becexp1}.
Due to its topological nature, the vortex thus created has the winding number $2F$, where $F$ is the hyperfine spin of the BEC.

A trapped BEC is unstable due to atom loss and vanishing gap along the axis of the trap during time-evolution of vortex creation, which necessitates a short creation time. CD driving of a single vortex creation is analyzed in \cite{beccd1}, which shows the CD
term $H_1$ corresponds to an unphysical magnetic field
${\bs B}_\mathrm{CD}$, which
does not satisfy $\mathrm{div}{\bs B}_\mathrm{CD}=0$ throughout the condensate. Accordingly, we impose 
Gauss's law only along a ring of a constant radius.
In spite of this approximation, it is shown that $H_1$ accelerates creation of a vortex and averts atom loss.

Furthermore, by taking advantage of short creation time, vortex pumping is possible to create vortices of a large winding number \cite{beccd2}. Solution of the Gross-Pitaevskii equation shows that pumping of vorticities with 20 cycles is possible
with reasonable parameter choices. Fast pumping of vortices
is also favorable to prevent a vortex with a large winding number from splitting into many vortices with small winding numbers.
 
Imperfect CD control is not always a bad thing. It is possible by taking advantage of this to create a topological link of the nematic vector $\bs d$ of an oblate $F=1$ BEC in the polar phase \cite{beccd3}. The condition $\mathrm{div}{\bs B}_\mathrm{CD}=0$ is imposed
only along a ring in the condensate. If the spins along the ring are rotated by $\pi$, spins in the other part of the condensate have different time-evolution from those along the ring. The resulting
structure is classified by the homotopy group $\pi_3(G)\simeq
\mathbb{Z}$, where $G=[U(1) \times S^2]/\mathbb{Z}_2$
is the order parameter space of the polar phase. The integer
of the homotopy group is called the Hopf charge, which is controllable by the radius of the ring. 
The polar phase is unstable
against decay into a more stable ferromagnetic phase.
The CD driving is necessary to create the link before the 
phase transition takes place.

\subsection{Fermi Gas}

STA of a Fermi gas is analyzed for both non-interacting and
strongly interacting cases in the unitary limit within the CD formalism, where the interaction strength is controlled by the Feshbach resonance technique \cite{Fg}. 
Superadiabatic expansion and compression of a Fermi gas in the unitary limit has been experimentally demonstrated.
The dynamics of a Fermi gas at high temperature is also studied,
where the Fermi gas is described by viscous hydrodynamics.

\subsection{Quantum Heat Engine and Quantum Refrigerator}

Maximal efficiency of heat engines may be attained if 
adiabatic process is achieved, which means the process is
slow. However, this implies the output power vanishes in the
adiabatic limit. STA may be applied to heat engines to
execute adiabatic processes in finite times so that the output
power is made finite. A typical example of a quantum heat engine
is the Otto cycle with a harmonic oscillator with variable trap
frequency as working medium, in which isentropic compression
and expansion are accelerated by STA \cite{ref:heat,ottreview}. 

It is proposed to use many-body systems to further enhance
the output power along with STA including the CD driving and the local CD driving \cite{hen1}.
This proposal has been experimentally demonstrated 
\cite{hen2}. Many-body quantum heat engine is also proposed in \cite{hen3}.

CD driving can be also employed to speed up and
enhance the efficiency and power of a quantum refrigerator.
\cite{qref} proposed to use a transmon superconducting
qubit coupled to two heat baths made of resonant circuits
to implement a quantum Otto refrigerator and evaluated
the heat fluxes, cooling power and thermodynamic efficiency
among others by using the GKLS equation.  

See \cite{mbreview} for a review on the many-body quantum thermal machines.

\subsection{Adiabatic Quantum Computing and Quantum Annealing}

Let $H_f$ be a diagonal Hamiltonian, for which we want to find the ground state. This seemingly trivial task is highly nontrivial
if the dimension of the Hilbert space on which $H_f$ acts
is large, $2^{1000}$ for example. In quantum annealing (QA for short), one starts
with a Hamiltonian $H_i$, whose ground state is easily prepared,
and then switches $H_i$ by $H_f$ adiabatically
with a time-dependent Hamiltonian $H(t)=f(t) H_f + (1-f(t)) H_i$
with $f(0)=0$ and $f(T)=1$.
If the time-evolution is adiabatic, one will end up with the 
ground state of $H_f$. 

A class of Ising Hamiltonian under a transverse field has been analyzed as an example of QA. To speed up the operation, 
\cite{AQC1,AQC2,AQC3} introduced Trotterization of the time-evolution operator and employed the variational formalism of CD driving introduced in section 4 to derive the CD Hamiltonian $H_1(t)$.

STA of QA is also analyzed in DI formalism without Trotterization \cite{TakaQA}.

\subsection{STA and Classical Nonlinear Integrable Systems}

Solutions of a class of time-dependent Schr\"odinger equation
that is reduced to a classical nonlinear integrable system
are obtained by using the equivalence of the dynamical invariant
equation and the Lax equation \cite{Lax1}.
Exact CD term is obtained whenever the
corresponding Lax pair exists. 

The same authors formulated STA in classical mechanics.
They employed the dispersionless Korteweg-de Vries (KdV) hierarchy to derive the CD term of a reference Hamiltonian \cite{Lax2}. They used the Hamilton-Jacobi theory to
define the generalized action that is directly related to the
adiabatic invariant.

\subsection{Many-body Systems}

 
As the number of controllable qubits has increased in recent years, STA for many-body systems become more and more important in both fundamental physics and applications.

\cite{mb1} considered the time-evolution through a quantum critical point, at which the gap between the ground state and the first
excited state vanishes and adiabaticity breaks down.
The 1-d transverse-field Ising model 
is mapped by the Jordan-Wigner transformation to a set of independent Landau-Zener Hamiltonians, for which $H_1$ is known \cite{3.berry}. However $H_1$
in real space is highly nonlocal. They also proposed approximate
STA by truncating the CD terms so that they have
finite ranges. An alternative scheme to avoid nonlocal CD term by tailoring the form of the CD interactions 
is also proposed \cite{mb2}. \cite{mb3} proposed protocols which minimizes excitation production in a closed quantum critical system
driven out of equilibrium.
 
In some cases, it is possible to obtained a local $H_1$
for a many-body systems. \cite{Lax1} finds a local
$H_1$ for the Toda lattice by taking advantage of the Lax pairs for classical integrable systems.  

Approximate CD term derived from the adiabatic gauge 
potential introduced in section 4 has been employed in various many-body systems in \cite{ref:sp}. 
This approach has been further developped in \cite{mb4},
where the adiabatic gauge potential is expanded in terms of
nested commutators between $H_0$ and $\partial_{\lambda} H_0$
and engineered by a Floquet protocol.

QAOA (Quantum Alternating Operator Ansatz) is a well known variational quantum algorithms. A new algorithm
called CD-QAOA has been developed, inspired by
the CD driving, for quantum many-body systems \cite{mb5}.
CD-QAOA combines the strength of continuous and discrete
optimization into a unified control framework.
It is demonstrated 
that CD-QAOA can be employed to prepare many-body ground-state using unitary evolution.

\subsection{CD Driving in Non-Hermitian Systems}

STA for Hermitian Hamiltonians can be generalized to
non-Hermitian Hamiltonians in both CD formalism and DI formalism \cite{nonh}. A reference Hamiltonian $H_0(t)$ defines two eigenvalue equations $H_0(t)|n(t) \ket
= E_n(t) |n(t) \ket$ and $H_0^\dag(t)|\hat n(t) \ket
= E_n^*(t) |\hat n(t) \ket$, where $\bra \hat n(t)|m(t) \ket = \delta_{mn}$. Then it can be shown that the CD term associated
with $H_0(t)$ is
\begin{equation}
H_1(t) = i \hbar \sum_n\left[ |\partial_t n(t) \ket \bra \hat{n}(t)|
-\bra \hat{n}(t)|\partial_t n(t)\ket |n(t) \ket \bra \hat{n}(t)|\right].
\end{equation}

\subsection{Couterdiabatic Born-Oppenheimer Dynamics}

There can be coexisting slow and fast degrees of freedom
in a single quantum system. 
In the Born-Oppenheimer approximation (BOA for short),
the slow degrees of freedom are regarded as frozen,
i.e. their kinetic energy is dropped,
when the Schr\"odinger equation of the fast degrees of freedom
is solved \cite{BOA}. The energy eigenvalues and eigenfunctions of the fast degrees of freedom are functions of the slow coordinates and
the energy eigenvalue acts as a potential energy for the
slow degrees of freedom.

It is challenging to obtain the spectral decomposition of the original Hamiltonian to construct the exact CD Hamiltonian in such a complex system in general. However, simpler CD terms for both the slow and fast variables are obtained under BOA,
in which the fast and the slow CD terms 
are obtained separately and then combined to produce
the CD term of the full system \cite{DC}.
This method, called the counterdiabatic Born-Oppenheimer
approximation (CBOA), has been applied to coupled harmonic
oscillators and two charged particles.

\section{Summary}

Counterdiabatic formulation of STA has been introduced here to make this special issue self-contained. STA of 1- and 2-spin systems and a harmonic oscillator are analyzed in detail.
They are expected to serve as illuminating examples to clarify
this technique. Some results are re-derived by using the
adiabatic gauge potential, which gives a different viewpoint to
the CD driving.
The relation between the CD
formulation and the DI formulation has been
discussed. Other subjects related to the CD formalism, not included in the main text, are briefly introduced for this article to serve as a
mini-review and a guide to further reading. 

STA, including the CD formalism, are rapidly growing field of research. They have found applications not only in physics
but also in chemistry, biology and mechanical engineering among
others. The readers are encouraged to find applications of STA 
in their own fields of research.


\enlargethispage{20pt}




\competing{The author declares that he has no competing interests.}

\funding{This research is supported by JSPS Grants-in-Aid
for Scientific Research (Grant Number 20K03795).}

\ack{The author is grateful to Ken Funo, Shumpei Masuda
 and Kazutaka Takahashi for useful communications.}




\begin{thebibliography}{99}

\bibitem{rev1} Torrontegui E, \textit{et al.} 2013.
Shortcuts to Adiabaticity. \textit{Adv. At. Mol. Opt. Phys.} \textbf{62}, 117.

\bibitem{rev2} Gu\'ery-Odelin D, \textit{et al.} 2019.
Shortcuts to adiabaticity: Concepts, methods, and applications.
\textit{Rev. Mod. Phys.} \textbf{91}, 045001. 

\bibitem{rev3} Masuda S and Rice S 2016.
Controlling Quantum Dynamics with Assisted Adiabatic Processes.
\textit{Advances in Chemical Physics} \textbf{159}, 51.

\bibitem{rev4} del Campo and Kim K. 2019. Focus on Shortcuts to Adiabaticity. \textit{New J. Phys.} 
https://iopscience.iop.org/journal/1367-2630/page/Focus-on-Shortcuts-to-Adiabaticity

\bibitem{1.emmanouilidou} Emmanouilidou A, Zhao X-G, Ao P, Niu Q.
2000. Steering an eigenstate to a destination. 
\textit{Phys. Rev. Lett.} \textbf{85}, 1626.

\bibitem{2.demirplak}  
Demirplak M and Rice S A. 2003. Adiabatic population transfer with control fields. \textit{J. Chem. Phys.
A} \textbf{107}, 9937.

\bibitem{2a.demirplak} Demirplak M and Rice SA. 2005. 
Assisted adiabatic passage revisited. 
\textit{J. Phys. Chem. B} \textbf{109}, 6838.



\bibitem{3.berry} Berry MV. 2009. Transitionless quantum driving.
\textit{J. Phys. A5: Math. Theor.} \textbf{42}, 365303.

\bibitem{ref:deffner} Deffner S, Jarzynski CJ, del Campo A 2014.
Classical and Quantum Shortcuts to Adiabaticity for Scale-Invariant Driving. \textit{Phys. Rev. X} \textbf{4}, 021013.

\bibitem{ref:adolfo} del Campo A 2013. Shortcuts to Adiabaticity by Counterdiabatic Driving. \textit{Phys. Rev. Lett.} \textbf{111},
100502.

\bibitem{ref:jarzynski} Jarzynski C 2013. Generating shortcuts to adiabaticity in quantum and classical dynamics. \textit{Phys. Rev. A} \textbf{88}, 040101(R).

\bibitem{4.Born} Born M and Fock V. 1928. Beweis des Adiabatensatzes. \textit{Z. Phys.} \textbf{51}, 165.

\bibitem{5.Kato} Kato T. 1995 \textit{Perturbation Theory for Linear Operators} (Classics in Mathematics 132). Berlin, Germany: Springer.

\bibitem{ref:chen} Chen X, \textit{et al.} 2010.
Fast Optimal Frictionless Atom Cooling in Harmonic Traps: Shortcut to Adiabaticity.
\textit{Phys. Rev. Lett.} \textbf{104}, 063002.

\bibitem{lewis} Lewis HR and Riesenfeld WB 1969.
An Exact Quantum Theory of the Time-Dependent Harmonic Oscillator and of a Charged Particle in a Time-Dependent Electromagnetic Field. \textit{J. Math. Phys.} \textbf{10}, 1458.

\bibitem{berryphase} Berry MV. 1984. Quantal phase factors accompanying adiabatic changes. \textit{Proc. R. Soc. Lond. A Math. Phys. Sci.} \textbf{392}, 45.

\bibitem{ktaka} Takahashi K. 2013. Transitionless quantum driving for spin systems. \textit{Phys. Rev. E} \textbf{87}, 062117.

\bibitem{DRx} Demirplak M and Rice S A. 2008.
On the consistency, extremal, and global properties of counterdiabatic fields. 
\textit{J. Chem. Phys.} \textbf{129}, 154111.

\bibitem{ref:Ibanez} Ib\'a\~nez S, \textit{et al.} 2012. Multiple Schr\"odinger Pictures and Dynamics in Shortcuts to Adiabaticity. \textit{Phys. Rev. Lett.} \textbf{109}, 100403. 


\bibitem{ref:mugax} Muga JG \textit{et al.} 2010.
Transitionless quantum drivings for the harmonic oscillator.
\textit{J. Phys. B: At. Mol. Opt. Phys.} \textbf{43}, 085509.

\bibitem{ref:taka_unitary} 
Takahashi K. 2015. Unitary deformations of counterdiabatic driving.
\textit{Phys. Rev. A} \textbf{91}, 042115.


\bibitem{ref:sp} Sels D and Polkovnikov A. 2017.
Minimizing irreversible losses in quantum systems by
local counterdiabatic driving. \textit{Prof. Natl. Acad. Sci. USA}
\textbf{114}, E3909-E3916.


\bibitem{ref:spr} Kolodrubetz M, Sels D, Mehta P, and Polkovnikov A. 
2017. Geometry and non-adiabatic response in quantum and
classical systems. \textit{Phys. Rep.} \textbf{697}, 1.





\bibitem{equiv} Chen X, Torrontegui E, and Muga JG.
2011. Lewis-Riesenfeld invariants and transitionless quantum driving.
\textit{Phys. Rev. A} \textbf{83}, 062116.


\bibitem{ref:Sarandy} Sarandy MS and Lidar DA. 2005. Adiabatic approximation in open quantum systems.
\textit{Phys. Rev. A} \textbf{71}, 012331.


\bibitem{openq1}
Vacanti G \textit{et al.}
2014. Transitionless quantum driving in open quantum
systems. \textit{New J. Phys.} \textbf{16}, 053017.


\bibitem{openq2}
Alipour S \textit{et al.}
2020. Shortcuts to Adiabaticity in Driven Open Quantum Systems:
Balanced Gain and Loss and Non-Markovian Evolution.
\textit{Quantum} \textbf{4}, 336.

\bibitem{sahar} Alipour S \textit{et al.}
2022. Entropy-based formulation of thermodynamics in arbitrary quantum evolution. \textit{Phys. Rev. A} \textbf{105},
L040201.



\bibitem{qsl review}
Deffner S and Campbell S. 2017.
Quantum speed limits: from Heisenberg's
uncertainty principle to optimal quantum control.
\textit{J. Phys. A: Math. Theor.} \textbf{50}, 453001.

\bibitem{ref:Santos}
Santos AC and Sarandy MS. 2015. Superadiabatic Controlled Evolutions and Universal Quantum Computation. \textit{Sci. Rep.}
\textbf{5}, 15775.

\bibitem{qsl sta}
Campbell S and Deffner S. 2017. Trade-off between speed and cost in shortcuts to adiabaticity. \textit{Phys. Rev. Lett.} \textbf{118}, 100601.

\bibitem{cost}
Zheng Y, Campbell S, De Chiara G and Poletti D. 
2016. Cost of counterdiabatic driving and work output. \textit{Phys. Rev. A} \textbf{94}, 042132.




\bibitem{frictionless}
del Campo A \textit{et al.}
2018.
Friction-Free Quantum Machines
in \textit{Thermodynamics in the Quantum Regime:
Fundamental Aspects and New Directions} ed. by
Binder F, Correa LA, Gogolin C, Anders J and Adesso G.
Springer Nature Switzerland. 


\bibitem{funok}
Funo K \textit{et al.}
2017. Universal Work Fluctuations During Shortcuts to Adiabaticity by Counterdiabatic Driving. \textit{Phys. Rev. Lett.}
\textbf{118}, 100602.

\bibitem{funoexp}
Zhang Z \textit{et al.} 2018. Experimental demonstration of work fluctuations along a shortcut to adiabaticity with a superconducting Xmon qubit. \textit{New J. Phys.} \textbf{20}, 085001.


\bibitem{funox}
Funo K, Shiraishi N and Saito K. 2019. 
Speed limit for open quantum systems. \textit{New J. Phys.}
\textbf{21}, 013006.

\bibitem{bukov}
Bukov M, Sels D and Polkovnikov A. 2019. 
Geometric Speed Limit of Accessible Many-Body State Preparation.
\textit{Phys. Rev. X} \textbf{9}, 011034.


\bibitem{qsl measure}
del Campo A. 2021. Probing Quantum Speed Limits with Ultracold Gases. \textit{Phys. Rev. Lett.} \textbf{126}, 180603.


\bibitem{funo1}
Funo K, Lambert N and Nori F 2021. General Bound on the Performance of Counter-Diabatic Driving Acting on Dissipative Spin Systems. \textit{Phys. Rev. Lett.}
\textbf{127}, 150401.


\bibitem{LCtheory}
Li Y-C and Chen X. 2016. Shortcut to adiabatic population transfer in quantum three-level systems: Effective two-level problems and feasible counterdiabatic driving. \textit{Phys. Rev. A} \textbf{94}, 063411.

\bibitem{atomstirap}
Du Y-X \textit{et al.}
2016. Experimental realization of stimulated Raman shortcut-to-adiabatic passage with cold atoms.
\textit{Nature Communications} \textbf{7}, 12479.

\bibitem{sc1}
Veps\"al\"inen A, Danilin S, Paraoanu GS. 2019.
Superadiabatic population transfer in a three-level
superconducting circuit. \textit{Sci. Adv.} \textbf{5}, eaau5999.

\bibitem{sc1.1} Zheng W \textit{et al.} 2022. Optimal control of stimulated Raman adiabatic passage in a superconducting qudit.
\textit{NPJ Quantum Information} \textbf{8}, 9.


\bibitem{sc2}
Yin Z \textit{et al.}
2022. Shortcuts to adiabaticity for open systems in circuit
quantum electrodynamics. \textit{Nature Communications} \textbf{13}, 188.




\bibitem{iontrap}
An S, Dingshun L, del Campo A and Kim K. 2016.
Shortcuts to adiabaticity by counterdiabatic
driving for trapped-ion displacement in phase
space. \textit{Nature Communications} \textbf{7}, 12999.

\bibitem{takaqb}
Takahashi K. 2013. How fast and robust is the quantum adiabatic passage? \textit{J. Phys. A:
Math. Theor.} \textbf{46}, 315304.


\bibitem{becv1}
Nakahara M. \textit{et al.}
2000. A simple method to create a vortex in Bose-Einstein condensate of alkali atoms. \textit{Physica B: Condens. Matter}
\textbf{284-288}, 17.


\bibitem{becv2}
Isoshima T \textit{et al.} 2000.
Creation of a persistent current and vortex in a Bose-Einstein condensate of alkali-metal atoms.
\textit{Phys. Rev. A} \textbf{61}, 063610.

\bibitem{becv3}
M\"ott\"onen M \textit{et al.}
2002. Continuous creation of a vortex in a Bose-Einstein condensate with hyperfine spin $F = 2$.
\textit{J. Phys.: Condens. Matter} \textbf{14}, 13481.
 
 
\bibitem{becexp1}
Leanhardt AE \textit{et al.}
2002. Imprinting Vortices in a Bose-Einstein Condensate using Topological Phases. \textit{Phys. Rev. Lett.} \textbf{89}, 190403.


\bibitem{beccd1}
Masuda S. \textit{et al.}
2016. Fast control of topological vortex formation in Bose-Einstein condensates by counterdiabatic driving. \textit{Phys. Rev. A} \textbf{93}, 013626. 


\bibitem{beccd2}Ollikainen T, Masuda T, M\"ott\"onen M and Nakahara M.
2017. Counterdiabatic vortex pump in spinor Bose-Einstein condensates. \textit{Phys. Rev. A} \textbf{95}, 013615.



\bibitem{beccd3}
Ollikainen T, Masuda M, M\"ott\"onen M and Nakahara M.
2017.
Quantum knots in Bose-Einstein condensates created by counterdiabatic control. \textit{Phys. Rev. A} 
\textbf{96}, 063609.


\bibitem{Fg}
Diao P \textit{et al.}
2018. Shortcuts to adiabaticity in Fermi gases. \textit{New J. Phys.} \textbf{20}, 105004.



\bibitem{ref:heat}
del Campo A, Goold J and Paternostro M. 2014.
More bang for your buck: Super-adiabatic quantum engines.
\textit{Sci. Rep.} \textbf{4}, 6208.

\bibitem{ottreview}
Kosloff R and Rezek Y. 2017. The Quantum Harmonic Otto Cycle.
\textit{Entropy} \textbf{19}, 136.
  
\bibitem{hen1}
Beau M, Jaramillo J and del Campo A. 2016.
 Scaling-Up Quantum Heat Engines Efficiently via
Shortcuts to Adiabaticity. \textit{Entropy} \textbf{18}, 168.

\bibitem{hen2}
Deng D \textit{et al.}
2018. Superadiabatic quantum friction suppression in
finite-time thermodynamics. \textit{Sci. Adv.} \textbf{4}, eaar5909.


\bibitem{hen3}
Hartmann A, Mukherjee V, Niedenzu W and Lechner W. 2020.
Many-body quantum heat engines with shortcuts to adiabaticity.
\textit{Phys. Rev. Research} \textbf{2}, 023145.


\bibitem{qref}
Funo K \textit{et al.}
2019. Speeding up a quantum refrigerator via counterdiabatic driving.
\textit{Phys. Rev. B} \textbf{100}, 035407.


\bibitem{mbreview}
Mukherjee V and Divakaran U. 2021. Many-body quantum thermal machines. \textit{J. Phys.: Condens. Matter} \textbf{33}, 454001.



\bibitem{AQC1} 
Hegade NN \textit{et al.}
2021. Shortcuts to Adiabaticity in Digitized Adiabatic Quantum Computing. \textit{Phys. Rev.Applied}
\textbf{15}, 024038.

\bibitem{AQC2} Chandarana P \textit{et al.}
Solano E, del Campo A and Chen X. 2022.
Digitized-counterdiabatic quantum approximate optimization algorithm. \textit{Phys. Rev. Research} \textbf{4}, 013141.

\bibitem{AQC3} 
Hegade NN, Chen X and Solano E. 2022.
Digitized-Counterdiabatic Quantum Optimization.
arXiv: 2201.00790.


\bibitem{TakaQA} Takahashi K. 2017. Shortcuts to adiabaticity for quantum annealing. \textit{Phys. Rev. A} \textbf{95}, 012309.


\bibitem{Lax1} Okuyama M and Takahashi K. 2016. From Classical Nonlinear Integrable Systems to Quantum Shortcuts to Adiabaticity. \textit{Phys. Rev. Lett.} \textbf{117} 070401,

\bibitem{Lax2}
Okuyama M and Takahashi K. 2017. Quantum-Classical Correspondence of Shortcuts to Adiabaticity. \textit{J. Phys. Soc. Jpn.} \textbf{86}, 043002.


\bibitem{mb1}
del Campo A, Rams MM and Zurek WH. 2012. Assisted Finite-Rate Adiabatic Passage Across a Quantum Critical Point:
Exact Solution for the Quantum Ising Model. \textit{Phys. Rev. Lett.}  \textbf{109}, 115703.

\bibitem{mb2}
Saberi H, Opatrn\'y T, M{\o}lmer K and del Campo A. 2014.
Adiabatic tracking of quantum many-body dynamics.
\textit{Phys. Rev. A} \textbf{90}, 060301(R).

\bibitem{mb3}
del Campo A and Sengupta K. 2015. Controlling quantum critical dynamics of isolated systems. \textit{Eur. Phys. J. Special Topics} \textbf{224}, 189.


\bibitem{mb4}
Claeys PW, Pandey M, Sels D and Polkovnikov A. 2019.
Floquet-Engineering Counterdiabatic Protocols in Quantum Many-Body Systems \textit{Phys. Rev. Lett.} \textbf{123}, 090602.

\bibitem{mb5}
Yao J, Lin L and Bukov M. 2021. Reinforcement Learning for Many-Body Ground-State Preparation Inspired by Counterdiabatic Driving. \textit{Phys. Rev. X} \textbf{11}, 031070.

\bibitem{nonh} Ib\'a\~nez S \textit{et al.}
2011. Shortcuts to adiabaticity for non-Hermitian systems.
\textit{Phys. Rev. A} \textbf{84}, 023415.


\bibitem{BOA}
Born M and Oppenheimer R. 1924.
Zur Quantentheorie der Molekeln.
\textit{Ann. Phys., Lpz.} \textbf{74}, 1.
 
\bibitem{DC}
Duncan CW and del Campo A. 2018.
Shortcuts to adiabaticity assisted by
counterdiabatic Born-Oppenheimer dynamics.
\textit{New J. Phys.}
\textbf{20}, 085003.


\end{thebibliography}
\end{document}